\documentclass[12pt]{article}
\input{epsf.sty}

\def\fo{\hbox{{1}\kern-.25em\hbox{l}}}

\def\slashchar#1{\setbox0=\hbox{$#1$}           
   \dimen0=\wd0                                 
   \setbox1=\hbox{/} \dimen1=\wd1               
   \ifdim\dimen0>\dimen1                        
      \rlap{\hbox to \dimen0{\hfil/\hfil}}      
      #1                                        
   \else                                        
      \rlap{\hbox to \dimen1{\hfil$#1$\hfil}}   
      /                                         
   \fi}                                         %

\def\boxll{\partial_\lambda\partial^\lambda}

\def\hn{H^{(\vec n)}}

\def\hide#1{[hidden stuff]}

\def\beq{\begin{equation}}
\def\eeq{\end{equation}}
\def\eq{\end{equation}}
\def\to{\rightarrow}

\def\mEt{\mbox{${\hbox{$E$\kern-0.6em\lower-.1ex\hbox{/}}}_T$}\, } 

\def\hsm{h_{\rm sm}}

\def\bsg{\ifmmode B\to X_s\gamma\else $B\to X_s\gamma$\fi}
\def\bsll{\ifmmode B\to X_s\ell^+\ell^-\else $B\to X_s\ell^+\ell^-$\fi}
\def\bstt{\ifmmode B\to X_s\tau^+\tau^-\else $B\to X_s\tau^+\tau^-$\fi}
\def\shat{\ifmmode \hat{s}\else $\hat{s}$\fi}

\newcommand{\newc}{\newcommand}

\newc{\lcal}{\int {\cal L}dt}
 
\newc{\LSP}{{\chi^0_1}}
\newc{\stauR}{{\tilde \tau_R}}
\newc{\stau}{{\tilde \tau_1}}
\newc{\mstop}{m_{\tilde{t}}}
\newc{\mHpm}{m_{H^\pm}}
\newc{\gsim}{\lower.7ex\hbox{$\;\stackrel{\textstyle>}{\sim}\;$}}
\newc{\lsim}{\lower.7ex\hbox{$\;\stackrel{\textstyle<}{\sim}\;$}}
\newc{\ie}{{\it i.e.}}          
\newc{\etal}{{\it et al.}}
\newc{\eg}{{\it e.g.}}          
\newc{\kev}{\hbox{\rm\,keV}}            
\newc{\mev}{\hbox{\rm\,MeV}}            
\newc{\gev}{\hbox{\rm\,GeV}}            
\newc{\tev}{\hbox{\rm\,TeV}}
\newc{\xpb}{\hbox{\rm\, pb}}
\newc{\xfb}{\hbox{\rm\, fb}}

\def\order#1{{\cal O}(#1)}
\newc{\mtop}{m_t}
\newc{\mbot}{m_b}
\newc{\mz}{m_Z}
\newc{\mw}{M_W}
\newc{\alphasmz}{\alpha_s(m_Z^2)}
\newc{\swsq}{\sin^2\theta_W}
\newc{\tw}{\tan\theta_W}
\newc{\cw}{\cos\theta_W}
\newc{\sw}{\sin\theta_W}
\newc{\BR}{\hbox{\rm BR}}
\newc{\zbb}{Z\to b\bar}
\newc{\Gb}{\Gamma (Z\to b\bar b)}
\newc{\Gh}{\Gamma (Z\to \hbox{\rm hadrons})}
\newc{\rbsm}{R_b^\hbox{\rm sm}}
\newc{\rbsusy}{R_b^\hbox{\rm susy}}
\newc{\drb}{\delta R_b}

\newc{\sgn}{\mbox{sgn}}
\newc{\tbeta}{\tan\beta}
\newc{\uL}{{\tilde u_L}}
\newc{\uR}{{\tilde u_R}}
\newc{\cL}{{\tilde c_L}}
\newc{\cR}{{\tilde c_R}}
\newc{\tL}{{\tilde t_L}}
\newc{\tR}{{\tilde t_R}}
\newc{\dL}{{\tilde d_L}}
\newc{\dR}{{\tilde d_R}}
\newc{\sL}{{\tilde s_L}}
\newc{\sR}{{\tilde s_R}}
\newc{\bL}{{\tilde b_L}}
\newc{\bR}{{\tilde b_R}}
\newc{\eL}{{\tilde e_L}}
\newc{\eR}{{\tilde e_R}}
\newc{\mhp}{m_{H^\pm}}
\newc{\mhalf}{m_{1/2}}
\newc{\emt}{{e/\mu /\tau}}

\newc{\lR}{\tilde{l}_R}
\newc{\lL}{\tilde{l}_L}
\newc{\nL}{\tilde{\nu}_L}
\newc{\na}{\chi^0_1}
\newc{\nb}{\chi^0_2}
\newc{\nc}{\chi^0_3}
\newc{\nd}{\chi^0_4}
\newc{\ca}{\chi^{\pm}_1}
\newc{\cb}{\chi^{\pm}_2}
\newc{\camp}{\chi^\mp_1}
\newc{\cbmp}{\chi^\mp_1}
\newc{\capos}{\chi^{+}_1}
\newc{\caneg}{\chi^{-}_1}
\newc{\phit}{\phi_t}
\newc{\phib}{\varphi_b}
\newc{\phiew}{\phi_{ew}}
\newc{\htz}{h^0_t}
\newc{\hbz}{h^0_b}
\newc{\hewz}{h^0_{ew}}
\newc{\hsmz}{h^0_{sm}}
\newc{\huz}{h^0_u}
\newc{\hsusyz}{h^0_{susy}}

\newc{\C}{{\cal C}}

\newcommand{\drawsquare}[2]{\hbox{%
\rule{#2pt}{#1pt}\hskip-#2pt
\rule{#1pt}{#2pt}\hskip-#1pt
\rule[#1pt]{#1pt}{#2pt}}\rule[#1pt]{#2pt}{#2pt}\hskip-#2pt
\rule{#2pt}{#1pt}}

\newc{\Dal}{\drawsquare{7}{0.6}}

\def\dofig#1#2{\epsfxsize=#1\centerline{\epsfbox{#2}}}
\def\dofigs#1#2#3{\centerline{\epsfxsize=#1\epsfbox{#2}%
   \hfil\epsfxsize=#1\epsfbox{#3}}}

\def\beq{\begin{equation}}
\def\eeq{\end{equation}}
\def\bea{\begin{eqnarray}}
\def\eea{\end{eqnarray}}
\catcode`@=11
\long\def\@caption#1[#2]#3{\par\addcontentsline{\csname
  ext@#1\endcsname}{#1}{\protect\numberline{\csname
  the#1\endcsname}{\ignorespaces #2}}\begingroup
    \small
    \@parboxrestore
    \@makecaption{\csname fnum@#1\endcsname}{\ignorespaces #3}\par
  \endgroup}
\catcode`@=12

\begin{document}
\begin{titlepage}

\begin{flushright}
CERN-TH/2000-051 \\
SNS-PH/2000-03 \\
UCD-2000-7 \\
LBNL-45201 \\
\end{flushright}






\huge
\bigskip
\bigskip
\begin{center}
{\Large\bf
Graviscalars from higher-dimensional \\ 
metrics and curvature-Higgs mixing}
\end{center}

\large

\vspace{.15in}
\begin{center}

Gian F.~Giudice$^a$, Riccardo Rattazzi$^b$, James D.~Wells$^c$

\small

\vspace{.1in}
{\it $^{(a)}$CERN Theory Division, CH-1211 Geneva 23, Switzerland \\}
\vspace{0.1cm}
{\it $^{(b)}$INFN and Scuola Normale Superiore, I-56100 Pisa, Italy \\ }
\vspace{0.1cm}
{\it $^{(c)}$Physics Department, University of California, 
      Davis, CA 95616, USA and \\
 Lawrence Berkeley National Laboratory, Berkeley, CA 94720, USA }

\end{center}
 
 
\vspace{0.15in}
 
\begin{abstract}

We investigate the properties of scalar fields arising from gravity
propagating in extra dimensions.  
In the scenario of large extra dimensions, proposed by Arkani-Hamed, Dimopoulos
and Dvali, graviscalar Kaluza-Klein 
excitations are less important than the spin-2 counterparts
in most processes. However, there are important exceptions.
The Higgs boson can mix to these particles by coupling to the
Ricci scalar. Because of the large number of states involved, this 
mixing leads, in practice, to a sizeable invisible width for the Higgs. 
In the
Randall-Sundrum scenario, the only 
graviscalar is the radion. It can be
produced copiously at hadron colliders by virtue of its enhanced 
coupling to two gluons through the trace anomaly of QCD.
We study both the production and decay of the radion, and compare
it to the Standard Model Higgs boson.  Furthermore, 
we find that radion detectability depends 
crucially on the curvature-Higgs boson mixing parameter.

\end{abstract}

\medskip

\begin{flushleft}
hep-ph/0002178 \\
February 2000
\end{flushleft}

\end{titlepage}

\baselineskip=18pt
\setcounter{footnote}{1}
\setcounter{page}{2}
\setcounter{figure}{0}
\setcounter{table}{0}


%

\tableofcontents

\section{ Introduction}

Recently it was recognized that the fundamental scale of quantum
gravity could be dramatically lower than the Planck scale provided
the Standard Model (SM) fields (gauge bosons and matter) propagate on
a 3-dimensional brane and gravity propagates in extra space dimensions.
The smallness of Newton's constant can then be explained by the
large size of the volume of compactification, as suggested
by Arkani-Hamed, Dimopoulos and Dvali (ADD)~\cite{Arkani-Hamed:1998rs,
largestring}. 
Another possibility,
proposed by Randall and Sundrum (RS)~\cite{rs}, is to have a non-factorizable
geometry where the 4-dimensional massless graviton wave function 
is localized away from our brane. If, as motivated by the hierarchy
problem, the fundamental gravity scale were not much bigger than a TeV,
these scenarios would have distinctive signatures in collider experiments.
In the ADD scenario, the production of an essentially continuum
spectrum of Kaluza-Klein (KK) graviton excitations gives rise to
characteristic missing energy 
signals~\cite{Giudice:1998ck,Mirabelli:1999rt,Han:1998sg,stringy,stringold}, 
while in the RS case one 
expects to see widely separated and narrow $J=2$ graviton modes
\cite{Davoudiasl:1999jd}.
Up to now phenomenological analyses have mainly focused on the
production of $J=2$ KK modes, although some studies of the $J=0$ modes have
been presented in 
refs.~\cite{Han:1998sg,Csaki:1999mp,Goldberger:1999un,
Grzadkowski:1999xh,Mahanta:2000zp}.
(The $J=1$  graviphotons
are not coupled at leading order to SM particles).
The purpose of this paper is to discuss in detail a few interesting aspects 
of the phenomenology of graviscalars.

At lowest order, the coupling to gravity waves $h_{AB}$ is 
\beq
{\cal L} = -{1\over 2} h_{AB} T^{AB} ~~~~A=(\mu ,i)~~~\mu=0,\dots ,3 ~~~
i=4,\dots ,D-1,
\label{couple}
\eeq
where $A,B$ run from 0 to $D-1$, while our brane world volume is along
$A=\mu$. We expect SM particles to correspond to brane excitations
for which the brane itself does not oscillate in the extra dimensions.
This means
that for processes involving SM particles, $T^{AB}$ has non-zero
components only along $A,B=\mu ,\nu$.
Therefore there is no ``direct'' coupling to the
scalar fields $h_{ij}$, $i,j>3$. However a coupling arises ``indirectly''
since the scalar $h_\mu^\mu$ mixes with the $h_{ij}$ in the
gravitational kinetic terms. The final result is that $h_{\mu \nu}$
at the brane location
includes graviscalars $\varphi_n$ when expanded in KK eigenmodes
\beq
h_{\mu\nu}= \sum_n
c_n \varphi^{(n)} \eta_{\mu\nu} + \left \{ J=2~ {\rm modes}\right \}.
\eeq
Here $c_n$ are effective coupling constants determined by the 
fundamental gravity scale and by the geometry of compactification.
Notice that the general covariance of the full theory constrains these
scalars to couple to the trace of the energy-momentum tensor.
One such scalar mode is the volume modulus
or radion $\varphi^{(0)}$, 
representing the fluctuations of the compactification
volume. Its coupling is $c_0=1/M_P$, where $M_P$ is the ordinary Planck mass
in the flat geometry of ADD,
while in the RS scenario it is roughly $c_0=1/(\Omega M_P)\sim 1/M_{\rm weak}$,
where $\Omega$ is the warp factor. On the other hand,
in the models that have been studied so far, the radion is a massless
mode and must be stabilized by some mechanism. Therefore the value of the 
radion mass depends on additional model-building assumptions
\cite{Sundrum:1999ns,Arkani-Hamed:1998kx,Goldberger:1999uk}, 
not just on the geometry.

In the case of 1 extra dimension ($D=5$) the radion is the only scalar 
mode. One could say that the massive KK excitations of the radion, along
with those of the graviphoton, are eaten by the J=2 modes to become massive
(a massive $J=2$ graviton has 5 helicity components the same as a massless set 
of fields $J=0\oplus 1\oplus 2$). For $D>5$ there is, however, a tower of 
massive  KK graviscalars. At each KK level $n $ there is one and only 
one such scalar $H^{(n)}$ coupled to $T_\mu ^\mu$~\cite{Giudice:1998ck}. 
In the flat ADD case
all these modes have $c_n\sim 1/M_P$.

One reason why previous studies have neglected 
graviscalars is that
$T_\mu^\mu$  vanishes at tree level for massless
fermions and massless gauge bosons, which are the quanta colliding in
high-energy experiments. For these particles the coupling arises
only at 1-loop via the trace anomaly. However things are different,
and more interesting, when there are also scalars propagating on the brane.
This is because already at the two-derivative level, scalars can couple
non-minimally to gravity. This corresponds to the fact that for a scalar
$\phi$ one can include in the four-dimensional effective
action terms involving the Ricci scalar $R(g)$ of the form
\beq
{\cal L}=MR(g)\phi -\frac{\xi}{2} R(g)\phi^2 +\cdots.
\label{terms}
\eeq
As is well known, even for a massless scalar, in order to have $T_\mu^\mu
=0$ at tree level one must add the above terms with $M=0$, $\xi=1/6$.
For all other choices, a massless scalar would couple to 
scalar gravitons already at tree level. 
In the case of the Standard Model
Higgs doublet, the gauge symmetry requires $M=0$, but 
it is difficult to anticipate the precise expected value for 
the numerical coefficient $\xi$.
Naive dimensional analysis suggests
$\xi$ to be of order unity. However,
if $\phi$ is a Goldstone boson, transforming non-linearly under a symmetry,
then  $M=0$, $\xi=0$. Therefore, if the Higgs is an approximate Goldstone
boson, then we expect a small value of $\xi$.

The most interesting 
consequence of the terms in eq.~(\ref{terms}) is a
kinetic mixing between the Higgs and the graviscalar. 
In the ADD case there is a huge number ${\cal O} (M_P^2)$ of graviton modes,
forming a near-continuum sufficiently degenerate in mass
with the Higgs boson to allow oscillations.
 We will show that such oscillations of the Higgs boson
into these scalars are equivalent to an invisible decay 
on collider time scales.
This is a generic prediction of theories with large extra dimensions:
no quantum number forbids
the mixing between the Higgs boson, which lives on the brane, 
and graviscalars in the bulk.
This invisible decay
 is also quantitatively interesting since, for $m_h< 2 m_W$, it 
competes with
the small visible Higgs decay width into $b \bar b$.

This paper is organized as follows. In section 2 we concentrate on the
large extra dimension case. We recall the properties 
of the KK scalar gravitons coupled to $T_\mu^\mu$ and discuss the implications
of the operator $\xi R\phi\phi^\dagger$. 
We focus on Higgs-graviton mixing, clarifying why it leads to 
invisible Higgs decays.
We discuss how this invisible channel can compete with the standard visible
ones. In section 3 we study the radion phenomenology in the RS scenario,
focusing on radion production and decay. In particular we point out the role of
the QCD trace anomaly in radion production via gluon fusion.

\section{Graviscalars from large extra dimensions}

Let us first 
briefly introduce our notation and recall previously known  results.
In this section we are considering the case in which
the full space-time is $M_4\times V_\delta$, where $V_\delta$ is the 
large-volume compactified space with $\delta\equiv D-4$ extra dimensions.
The action of the theory is given by the $D$-dimensional Einstein term
plus a 4-dimensional brane term
\beq
S= \frac{\bar M_D^{2+\delta}}{2}\int d^Dx \sqrt{-g}R
    +\int d^4x \sqrt{-g_{\rm ind}}{\cal L_{\rm sm}}.
\label{action1}
\eeq
Here $\bar M_D$
is the $D$-dimensional reduced Planck constant and $g_{\rm ind}$
is the induced metric on the brane. For the case of a flat brane
we are considering we have simply $(g_{\rm ind})_{\mu \nu} =
g_{\mu\nu}$ for $\mu,\nu=$ 0 to 3. 
Upon integrating eq.~(\ref{action1})
over the extra-dimensions we obtain the 4-dimensional
reduced Planck mass $\bar M_P$ \cite{Arkani-Hamed:1998rs}
\beq
\frac{1}{8\pi G_N}\equiv
\bar M_P^2=\bar M_D^{2+\delta} V_\delta =\bar M_D^{2+\delta} (2\pi r)^\delta ,
\eeq
where we have assumed for simplicity that the compactified space with
volume $V_\delta$ is a $\delta$-torus of radius $r$.
If we define for convenience $M_D=(2\pi)^{\delta /(2+\delta )} 
\bar M_D$ then we  
can identify
\beq
M_D^{2+\delta}=\bar M_P^2/ r^{\delta}.
\label{MDeq}
\eeq
If $r^{-1}\ll \bar M_P$ then it is possible to have
$M_D\ll \bar M_P$, perhaps as low as the weak scale.

By linearizing
$g_{AB} = \eta_{AB}+h_{AB}$ and expanding $h_{AB}$
in Fourier modes
\beq
h_{AB}=\sum_{n_1= -\infty}^{\infty} \cdots \sum_{n_\delta = -\infty}^{\infty}
   \frac{h^{(n)}_{AB}(x)}{\sqrt{V_\delta}} e^{i n^j y_j/r},
\label{fourier}
\eeq
one finds the physical KK modes \cite{Giudice:1998ck}. In eq.~(\ref{fourier})
$y$ are the coordinates of the extra dimensions, and $y=0$ defines the location
of our brane.
Here we are only interested in the scalar modes living in $h_{\mu\nu}$.
(The phenomenology of scalar modes corresponding to brane deformations
has been recently considered in ref.~\cite{DeRujula:2000he}.)
As shown in ref.~\cite{Giudice:1998ck},  at each KK level $\vec n$ there is
only one such mode $H^{(\vec n)}$ which, after proper normalization, has a 
lagrangian 
\bea
{\cal L} & =&\sum_{{\rm all}~\vec n} \left[
-\frac{1}{2} H^{(-\vec n)} (\boxll +m_n^2) H^{(\vec n)} 
+\frac{\kappa}{3\bar M_P}
     H^{(\vec n)}
T_{\mu}^{ \mu}\right]
\cr
\kappa &\equiv& 
\sqrt{\frac{3(\delta-1)}{\delta +2}}~~~~m_n^2\equiv \frac{\vec n^2}
{r^2}.
\label{freelag}
\eea
Notice that the equation above is valid only for $\delta > 1$.
For $\delta =1$ there are no physical propagating KK graviscalars.

\subsection{Scalar-curvature term}

The SM energy-momentum tensor is defined as 
$T^{\mu\nu}=-\eta^{\mu\nu}{\cal L}_{\rm sm}
+2\delta {\cal L}_{\rm sm}/\delta g_{\mu\nu}$.
However, 
already at the two-derivative level we can add to ${\cal L}_{\rm sm}$
a term coupling the Higgs field $\phi$ to the Ricci scalar
of the induced 4-dimensional metric 
\beq
S = -\xi\int d^4x \sqrt{-g_{\rm ind}}R(g_{\rm ind})\phi^\dagger \phi .
\label{lcs}
\eeq
This term determines an additional effective contribution to $T_{\mu\nu}$
as can be seen by expanding 
$\sqrt{-g}R$ in
the weak gravitational field $g_{\mu\nu}=\eta_{\mu\nu}+h_{\mu\nu}$
\beq
\sqrt{-g}R = (\boxll \eta^{\mu\nu}-\partial^\mu\partial^\nu )
                 h_{\mu\nu}(x,y=0)
\eeq
and taking the variation of eq.~(\ref{lcs}) with respect to $h_{\mu\nu}$
\beq
T^{({\rm new})}_{\mu\nu}=
   T^{({\rm naive})}_{\mu\nu}+2\xi (\eta_{\mu\nu}\boxll 
     -\partial_\mu\partial_\nu) (\phi^\dagger \phi).
\label{new tensor}
\eeq
The added term is a total derivative and is automatically conserved.
The addition of this term to the stress tensor is a standard technicality
in field theory in order to ``improve'' the properties of the dilatation 
current $S_\mu=T_{\mu\nu} x^\nu$. In our case, however, since gravity
is coupled, the presence of this term has physical consequences.
The added term represents a spin-0 field, so on-shell the coupling of
the spin-2 graviton KK modes are
not affected. The coupling to
the scalar KK gravitons $H^{(\vec n)}$ is crucially changed
by the addition $\Delta T_\mu^\mu=6\xi \Dal (\phi\phi^\dagger )$.

It is clarifying to write ${T^{({\rm new})}}^\mu_\mu$ in the SM by working in 
the 
unitary gauge. Here $\phi$ reduces to the physical real Higgs
field $h$, according to $\phi=[(
v+h)/{\sqrt 2},0]$, with $v=246$~GeV.
Then, after using the SM equations of motion, we have
\bea
{T^{(\rm new )}}^\mu_\mu &=& -(1-6\xi)\left \{ \partial_\mu h\partial^\mu
 h + M_V^2 V_{A\mu}V_A^\mu \left( 1+\frac{h}{v}\right)^2 -
m_{ij}\bar\psi_i
\psi_j \left( 1+\frac{h}{v}\right) \right. \nonumber \\
 &&\left. -\frac{\lambda}{2}\left( v+h\right)^4
\right \} -(1-3\xi)~m_h^2~(v+h)^2 .
\label{newtrace}
\eea
Here $V_{A\mu}$ and $\psi_i$ are respectively the massive vector bosons and 
fermions, and we have taken a Higgs potential $V=(\lambda |\phi|^4
-m_h^2 |\phi|^2)/2$. 
Notice that we recovered the standard result that for conformal
coupling $\xi=1/6$ the trace is proportional to the truly conformal
breaking parameter, the Higgs mass $m_h^2=\lambda v^2$.
However for $\xi\not = 1/6$, the coupling to
the scalar gravitons $H^{(\vec n)}$ is not suppressed by the Higgs mass.
This is manifest for the kinetic and quartic terms
of $h$ in eq.~(\ref{newtrace}). It is also true, but less evident,
 for the massive vector bosons whose coupling is proportional to 
$M_V^2\propto m_h^2$. This is because the  wave function for a
longitudinally polarized vector at high energy
is proportional to the term $p^\mu/M_V$, 
which removes the mass suppression in eq.~(\ref{newtrace}).
This result is an obvious consequence of the equivalence theorem, by which the
vector boson term in eq.~(\ref{newtrace}) can be replaced with the Goldstone
boson kinetic term at zeroth order in $g_{\rm weak}$. 
The results of ref.~\cite{Han:1998sg} for the couplings of $H^{(\vec
n)}$ correspond
to the choice $\xi=0$. This explains the absence of a $m_h^2$ (or $M_V^2$)
suppression in the calculated scalar graviton width. However, there is really 
no good argument to prefer $\xi=0$. 
For instance, quantum corrections renormalize
 $\xi$. So it is reasonable to expect $\xi$ to be of order 1. This will
be our assumption in the rest of the paper.

A crucial feature of eq.~(\ref{newtrace}) is a linear term in the
Higgs field $-6\xi v m_h^2 h$, which is purely generated by
the interaction in 
eq.~(\ref{lcs}). This term can be read directly from eq.~(\ref{new tensor}),
by noticing that $T^{({\rm naive})\mu\nu}=-2\delta {\cal L}_{\rm sm}/\delta 
g_{\mu\nu}$
is at least quadratic in $h$ on the stationary point of ${\cal L}_{\rm sm}$
and by using the lowest order equation of motion $\Dal h=-m_h^2 h$.
{}Because of the interaction 
in eq.~(\ref{freelag}),  this linear term leads to an
$H^{(\vec n)}$-Higgs mass mixing
\beq
{\cal L}_{\rm mix} = - m^2_{\rm mix} h \sum_{\vec n} H^{(\vec n)},~~~~
m^2_{\rm mix}\equiv\frac{2\kappa \xi v m_h^2}{\bar M_P} .
\label{mix}
\eeq
In the presence of such mixing, to extract physical information,
one would normally try to diagonalize the mass matrix to find the
eigenvalues and mixing angles. Here this task is made difficult
by the huge number $\order{M_P^2}$
 of graviton levels mixing with $h$.
(We recall that the level density around a KK mass $m_G$ is $dN=V_\delta
m_G^{\delta-1}d m_G/(2\pi)^\delta$.)
Moreover some of them lie very close to 
$m_h^2$ so that perturbation theory would not work. On the other hand, we
also do not need all the information contained in the spectrum, as
we will never be able to observe a single graviton mode. 
The information we need can be synthesized into a single
expression for the tree level Higgs propagator $G_h(p^2)$
\beq
\langle h h\rangle_{p^2}\equiv G_h(p^2)=\sum_{a} \frac{|U_a|^2}{p^2-m_a^2+i\epsilon},
\label{eigenprop}
\eeq
where $U_a$, $m_a$ are the mixing angles and eigenmasses. We can 
implicitly write $G_h$ by formally inverting the
mass matrix or, which is the same, by summing up  the
$m_{\rm mix}^2$ insertions. This gives
\beq
G_h(p^2) = \frac{i}{p^2-m^2_h+\Sigma (p^2)+i\epsilon},
\label{implicit}
\eeq
\beq
\Sigma (p^2) \equiv -m^4_{\rm mix}\sum_{\vec n}
  \frac{1}{p^2-|\vec n/r|^2+i\epsilon}.
\label{discrete}
\eeq
Now, $\Sigma(p^2)$ is a badly-behaved function (in fact it is a distribution)
with singularities at each pole $\vec n^2/r^2$. However, as it is 
intuitively expected and as we will prove in the sect.~2.2,
for observables that do not resolve the single mode the
above discrete sum $\sum_{n}$ can be turned into a $d^\delta q_T$ integral.
Then the result becomes quite simple
\beq
\Sigma(p^2)\simeq -m_{\rm mix}^4\frac{\bar M_P^2}{M_D^{2+\delta}}\int
\frac{d^\delta q_T}{p^2-q_T^2+i\epsilon}=
\kappa^2\xi^2S_{\delta -1}
     \frac{v^2m_h^4}{M^{2+\delta}_D}
   \left[i\pi p^{\delta-2}+ F(\Lambda,p^2)\right].
\label{continuum}
\eeq
Here $S_{\delta -1}$ is the surface of a unit-radius sphere in $\delta$ 
dimensions and 
$F(\Lambda, p^2)$ is a cut-off dependent
real polynomial of degree $\delta-2$
and $\Lambda$ indicates the cut-off scale  at which the effective theory 
breaks down.  
$F(\Lambda, p^2)$, the real part of $\Sigma(p^2)$,
 corresponds to a set of local counterterms
renormalizing the Higgs pole mass, residue and other 
input parameters. While these renormalizations have physical consequences,
(for instance the wave function renormalization
can correct the on-shell higgs couplings by an amount ${\cal O}(m_h^4/M_D^4)$)
they will play no role in our discussion, as it will become clear in 
a moment.
This is in contrast to virtual
graviton mediated processes at high-energy colliders, 
where the value of this
function must be hypothesized in order to calculate a 
rate~\cite{Giudice:1998ck,Han:1998sg,Hewett:1999sn}.

As demonstrated above, the
function $\Sigma (p^2)$ has a {\it calculable} imaginary part.
This can be thought of as a decay of the Higgs into the scalar gravitons
$H^{(\vec n)}$ with partial width
\beq
\Gamma_{G}=\frac{1}{m_h}{\rm Im}\, [\Sigma(p^2=m^2_h)]
  =\pi\kappa^2\xi^2 v^2\frac{m_h^{1+\delta}}{M_D^{2+\delta}}S_{\delta -1}.
\label{inv width}
\eeq
This result is quite intuitive, when we think in the $V_\delta \to \infty$
limit. The Higgs excites bulk gravitons via eq.~(\ref{mix}),
but these escape
in the extra-dimension and never come back. So even if $h$ could not decay
into any SM particle 
there would still be no
asymptotic Higgs state on the brane because of decay into {\it just one}
graviton. 
This is possible because momentum in the direction transverse to the brane
is not conserved. This is analogous to the case of the radiative decay 
of an excited nucleus, in the large-mass limit.
In sect.~2.2 we will 
show that at finite $V_\delta$  this decay arises from the overlapping 
 oscillations of the Higgs
into the many accessible 4-dimensional $H^{(\vec n)}$ fields.   

Finally,
to include the effects of the usual decay modes of the SM Higgs boson
we should just add to $\Sigma(p^2)$
the corresponding one-loop bubble diagrams 
with the appropriate imaginary parts.
We will discuss the phenomenological
consequences in sect.~2.3.

\subsection{Higgs-graviton mixing: discrete versus continuum}

In this section we prove that the mixing to a large
number of closely spaced  levels is equivalent to a decay. 
The readers who are already convinced of this assertion and
are satisfied  with the arguments given in sect.~2.1 may prefer to
skip this section.

\begin{figure}[t]
\dofig{3.5in}{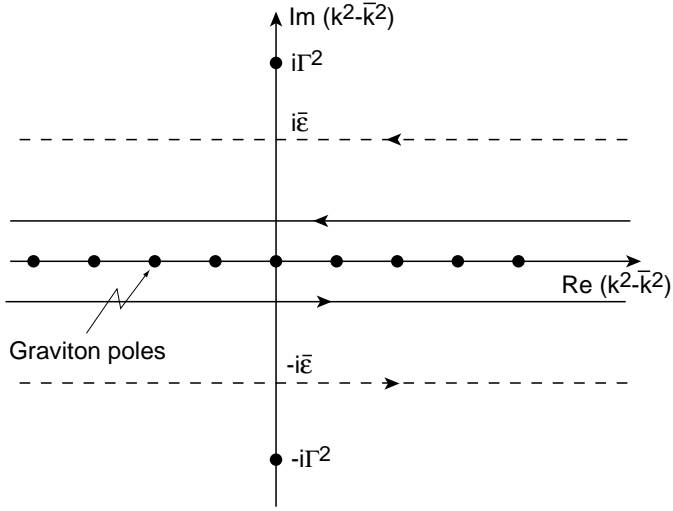}
\caption{The contour of integration ${\cal C}$ (solid line) and
$\C_{\bar \epsilon}$ (dashed line) in the $k^2-{\bar k}^2$
complex plane. It is assumed that 
$\Delta m_G^2\ll\bar\epsilon\ll\Gamma^2$, where $\Delta m_G^2$
is the splitting between graviton poles. \label{poles}}
\end{figure}

Consider, for the sake of argument, a close relative to the SM where
the Higgs boson is the lightest particle and is stable.
The signal of its production is missing energy.
For instance, the process $e^+e^-\to Z h$ leads to a missing momentum 
$k^2$ delta function at $k^2=m_h^2$. In the presence of the mixing in 
eq.~(\ref{mix}),
the missing momentum cross-section for the same reaction will become
\beq
\frac{d\sigma}{d k^2}=\sum_a |U_a|^2 \delta(k^2-m_a^2)\sigma_{SM}(m_a^2,E) .
\label{distribution}
\eeq
Here $\sigma_{SM}(m_a,E)$ is the SM cross section to produce a Higgs
of mass $m_a$ at center of mass energy $E$, and $d\sigma/dk^2$ is a highly
discontinuous function that we do not access experimentally when 
$V_\delta\to \infty$. What we measure experimentally is the convolution 
\beq
\int \frac{d\sigma}{d k^2} f(k^2)dk^2
\label{convo}
\eeq
where $f$ is any smooth test function spreading over a {\it finite}
$k^2$ range. For instance, we can imagine the functions $f$ to be of the form
\beq
f(k^2)=\frac{C}{((k^2-\bar k^2)^2+\Gamma^4)^n}
\eeq
where $\Gamma$ characterizes the width of $f$ around a point $\bar k^2$.
Equation~(\ref{convo}), by using eqs.~(\ref{distribution}) and 
(\ref{eigenprop}),
can be conveniently written as a contour integral
\beq
\int \frac{d\sigma}{d k^2} f(k^2)dk^2\,=\,
\oint_{\cal C} \sigma_{SM}(k^2,E)f(k^2)G_h(k^2)\frac{dk^2}{2\pi i}
\label{contour}
\eeq
where ${\cal C}$ consists of the two lines, one just above and the other just
below the real $k^2$ axis, oriented as in Fig.~\ref{poles}.
The path $\C$ can be moved away from the real axis, without affecting
the result, as long as the poles of $f$ at ${\rm Im}(k^2)=\pm \Gamma^2$
are not crossed.
In the limit $V_\delta\to \infty$ the level separation $\Delta m_G\sim
1/V_\delta$ becomes $\ll \Gamma$, the width of $f$. Now it is convenient
to evaluate eq.~(\ref{contour}) on a path $\C_{\bar \epsilon}$ defined by 
${\rm Im} k^2=\pm
\bar \epsilon$ with  $\Delta m_G^2\ll\bar\epsilon\ll\Gamma^2$. On 
$\C_{\bar\epsilon}$ we can still approximate $\sigma(k^2\pm i\bar \epsilon)
\simeq \sigma(k^2)$ and $f(k^2+i\bar\epsilon)\simeq f(k^2)$. Moreover,
since $\Delta m_G^2\ll \bar\epsilon$, the discrete sum 
defining $G_h$ in eqs.~(\ref{implicit}, \ref{discrete})
involves a function $(k^2-m_n^2+i\bar\epsilon)^{-1}$ which is no
longer rapidly varying from mode to mode. Therefore we can safely convert
the sum to an integral
\beq
{1\over V_\delta}\sum_{n}\frac{1}{k^2-(n/r)^2+i\bar\epsilon}=
\int \frac{d^\delta q_T}{k^2-q_T^2+i\bar\epsilon}+{\cal O}\left (
\frac{1}{V_\delta\bar \epsilon}\right )
\eeq
where $i \bar\epsilon$ plays now the role of the usual 
$i\epsilon$. We conclude that 
for $V_\delta\bar\epsilon\sim \bar\epsilon/\Delta m_G^2\to\infty$,
$\Sigma(p^2)$ can be replaced by the continuum limit 
in eq.~(\ref{continuum}) and the integral over $\C_{\bar\epsilon}$
 reduces to
\beq
\int \frac{d\sigma}{d k^2} f(k^2)dk^2\simeq \frac{1}{\pi} \int 
\sigma_{SM}(k^2,E)
f(k^2)\frac{m_h\Gamma_G}{(k^2-m_h^2)^2+m_h^2\Gamma_G^2}{dk^2}
\eeq
This is the usual Breit-Wigner formula for a resonance of width
$\Gamma_G$, where $\Gamma_G$ is given in eq.~(\ref{inv width}).
Technically what we have shown is that the distribution in 
eq.~(\ref{distribution}) when acting on a smooth function can be replaced with
a continuous Breit-Wigner function. The essential ingredient in the
simplification of the final expression is that it represents
an {\it inclusive} quantity over all the modes
that for $V_\delta\to \infty$ become degenerate with the Higgs.
On the other hand, {\it exclusive} quantities, representing the
production of individual modes, do not have a smooth $V_\delta\to\infty$ 
limit. 

The Higgs-graviton mixing we are considering should be contrasted to the
case of a mixing between the SM left-handed neutrinos and fermions living
in the bulk \cite{Arkani-Hamed:1998vp,Dienes:1999sb}. The 
resulting neutrino mass
is practically zero  (it is indeed a $1/M_P$ effect), so the KK level density
that it experiences is very small and only a few levels mix significantly.
So the appropriate description is in terms of usual mixing.
Indeed, also in our case the mixing looks like a decay only as
long as $\Gamma_G$ is much bigger than the  separation of KK levels,
{\it i.e.} as long as
 $m_h^\delta \gg M_D^{2+\delta}/(vM_P)$, see eq.~(\ref{inv width}).

It is instructive to present another point of view on the
Higgs-graviton mixing
based on time oscillations. Let us work in the non-relativistic
limit to simplify notation. Because of  eq.~(\ref{eigenprop}), we can write the
Higgs state as a sum over the mass eigenstates
\beq 
|h\rangle=\sum_a U_a|a\rangle.
\eeq
The $|h\rangle$ to $|h\rangle$ amplitude at time $t$ is
\beq
A(t)=\langle h(t)|h(0)\rangle=\sum_a e^{-i m_a t}|U_a|^2 .
\eeq
If we want to study the behavior of $A(t)$ for times $t\ll 1/\Delta m_G$,
for which the graviton levels cannot be resolved, then we should consider
the Fourier (Laplace) transform
\beq
\hat A(\omega)=\int_0^\infty e^{i(\omega +i\bar\epsilon)t} A(t)=\sum_a
\frac{1}{\omega-m_a+i\bar\epsilon} |U_a|^2
\label{nonrel}
\eeq
where $\bar\epsilon$ is taken to be much bigger than $\Delta m_G$ but much
smaller than any laboratory energy or inverse  time. As was the case
before, $(\omega-m_a+i\bar\epsilon)^{-1}$ is now a smooth function
of $m_a$, and we can safely convert the sum to an integral. Therefore, with
the proper normalization,
eq.~(\ref{nonrel}) reduces
to the non-relativistic limit of eq.~(\ref{implicit}), with $\Sigma$
given by the continuum limit eq.~(\ref{continuum})
\beq
\hat A(\omega)= \frac{1}{\omega-m_h+i\Gamma_G/2}.
\eeq
So even though at times $t>1/ \Delta m_G$, the amplitude $A(t)$ may display
a complicated oscillation pattern reflecting the structure of the graviton
spectrum, at times $\ll 1/\Delta m_G$ the various oscillation amplitudes
sum up to give an exponential decay $|A|\sim {\rm exp} (-t \Gamma_G/2)$.

\subsection{Invisible Higgs width}

The partial width $\Gamma_G$ contributes to the invisible width of
the Higgs since the $H^{(\vec n)}$ will not interact with the detector
or decay inside.  The Higgs boson also has visible decay modes 
in the SM that will compete with this invisible partial
width.  For Higgs bosons with mass below about $150\gev$ the total
width into SM states is less than $20\mev$.  This is
extremely narrow, and so any new decay modes may likely dominate
over SM modes.  Above the $h\to WW$ threshold, the
total width is unsuppressed and so new decay modes are not likely to
dominate, but can compete with SM states at best.

\begin{figure}[t]
\dofig{6in}{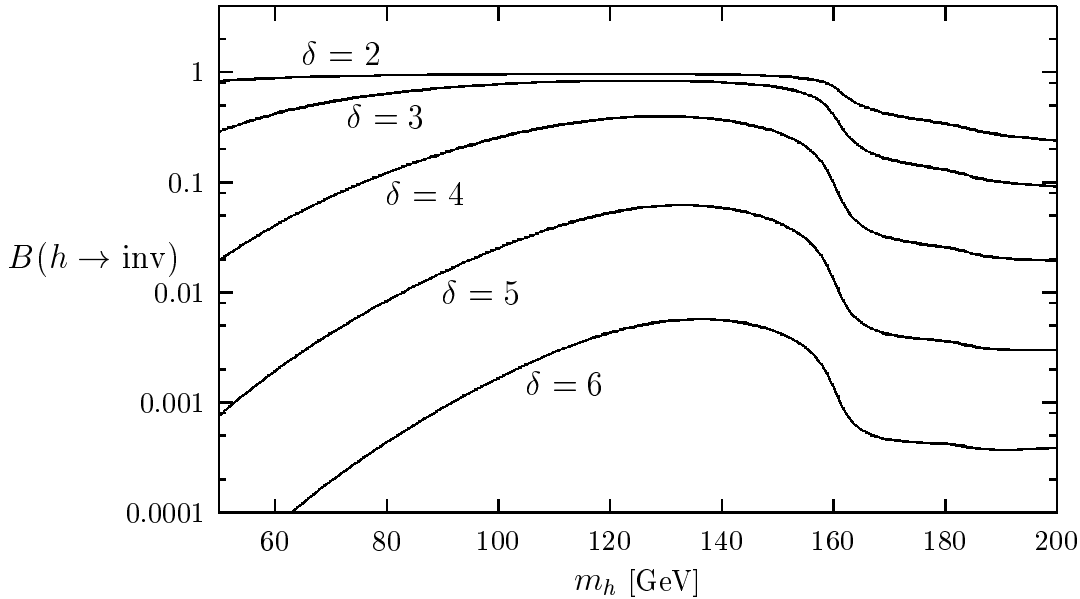}
\caption{Branching fraction of the Higgs boson to
decay invisibly as a function of its mass, for  $M_D=2\tev$ and
$\xi =1$. The rapid decrease at
$m_h\simeq 160\gev$ is due to the onset of $h\to WW$ on-shell decays.
\label{inv_Higgs}}
\end{figure}

In Fig.~\ref{inv_Higgs} we plot the branching fraction of the
Higgs boson to decay invisibly,
\beq
B(h\to {\rm invisible})=\frac{\Gamma_G}{\Gamma_{\rm SM}+\Gamma_G}.
\eeq
The plot is made for $M_D=2\tev$, $\xi=1$, and for various numbers of
extra dimensions $\delta$.  The scaling with respect to these parameters
can be obtained from eq.~(\ref{inv width}).

At LEP2, the invisible decaying Higgs boson has been searched for
by all four collaborations~\cite{invisible Higgs}.  
No evidence has been found for this
state, and the lower limit on the mass of a Higgs boson with SM production
cross
section and 100\% invisible branching ratio is now 98.9~GeV.
Future analyses with $\sqrt{s}\simeq 200\gev$ are expected to
either find the invisible Higgs boson or exclude its existence
up to nearly $105\gev$.

At the Tevatron with $30\xfb^{-1}$ of integrated luminosity, 
an invisibly decaying Higgs boson could be discovered
at greater than $5\sigma$ significance if its mass is below $110\gev$,
and it could be excluded at the 95\% C.L. if its mass is less than
$145\gev$~\cite{Martin:1999qf}.  
At the LHC, searches 
are likely to find evidence for such a Higgs boson if its mass is below about 
$250\gev$~\cite{lhc inv higgs}.

\begin{figure}[t]
\dofig{6in}{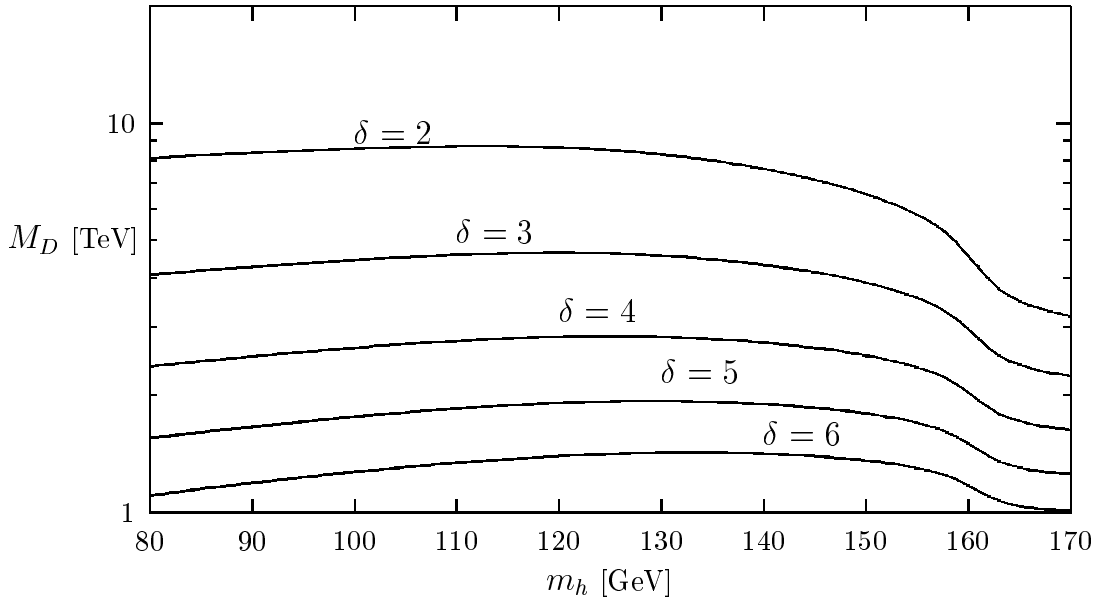}
\caption{Upper bound on the sensitivity 
to $M_D$ from invisible width measurements of
the Higgs boson.  We have approximated the width measurement 
capability as the expected uncertainty in the SM 
Higgs width determination~\cite{gunion talks} at a
muon collider (on-shell scan with $0.4\xfb^{-1}$) and
$e^+e^-$ linear collider ($\sqrt{s}=500\gev$ with $200\xfb^{-1}$)
given $\xi=1$ and various choices of the number of extra dimensions
$\delta$.  
\label{MD_muon}}
\end{figure}

It is conceivable that a future muon collider  working
around the Higgs resonance will be capable of measuring the 
Higgs total width $\Gamma^{\rm tot}_h$
to a precision better than 20\% in the mass
range $110\gev\lsim m_h\lsim 150\gev$~\cite{Casalbuoni:1999fx,Yello}. 
Combining high luminosity $e^+e^-$ collider data
($500\gev$ with $200\xfb^{-1}$)
with muon collider data (on-shell scan of $0.4\xfb^{-1}$), it
is reasonable to expect better than 12\% total width determination in the
mass range $80\gev< m_h < 170\gev$~\cite{gunion talks}.
We assume that the sensitivity to $\Gamma_G$ will be
close to the ability to determine the width, $\Delta \Gamma_h^{\rm tot}$.
This in turn implies that the sensitivity to $M_D$ is 
\beq
M_D < \left[ \frac{\pi\kappa^2\xi^2v^2m^{\delta+1}_hS_{\delta -1}} 
   {\Delta\Gamma_h^{\rm tot}}\right]^{\frac{1}{\delta+2}}.
\eeq
In Fig.~\ref{MD_muon} we plot these sensitivity limits,
using $\Delta\Gamma_h^{\rm tot}$ values from ref.~\cite{gunion talks}
as a function of $m_h$
for different numbers of extra dimensions, and for $\xi=1$.  One can
scale the result easily for other values of $\xi$ according to
the above equation.  We can see that the sensitivity to $M_D$ is best
for lower numbers of extra dimensions as is usually the case because the
phase space density of ``light'' KK modes is higher with fewer dimensions.

An invisibly decaying Higgs boson is not necessarily evidence for Higgs
boson oscillations into graviscalars.  There are many extensions of the
SM that predict the Higgs boson decaying into undetectable 
states (see {\it e.g.} ref.~\cite{Martin:1999qf}).
For example, Higgs decays into singlet scalars or Majorons are
just two four-dimensional examples of an invisible partial width.  Furthermore,
there are other intrinsically extra-dimensional decays beyond the decays to 
graviscalars.  If the right-handed neutrino
lives in the bulk, Higgs decays into all accessible neutrino pairs can have 
an ${\cal O}(1)$ branching 
fraction~\cite{Arkani-Hamed:1998vp,Martin:1999qf}.  Distinguishing between
these possibilities is extremely difficult without additional observables.
Nevertheless, the effects do cause deviations from SM expectations, and
so it is meaningful to speak of sensititivity to $M_D$
in these theories.  Although
dependent on unknown parameters such as $\xi$ and $\delta$, 
the multi-TeV sensitivity 
to $M_D$ that we have demonstrated in the analysis for next generation
colliders, such as the LHC and a muon collider,
compares favorably with sensitivities derived from calculable
external KK spin-2 graviton processes~\cite{Giudice:1998ck,Mirabelli:1999rt}.

\subsection{Direct graviscalar production at LEP2}

In the previous section we discussed effects of graviscalars
on the invisible width of the Higgs boson.  One can also search
for graviscalars in direct production at high-energy colliders.
Earlier we suggested that spin-0 KK effects were usually not as relevant to 
direct collider physics probes/constraints as the spin-2 KK effects,
because they couple only to the trace of the energy-momentum tensor.
At very high-energy scattering in hadron colliders, they can be
produced by gluon fusion, but their production is suppressed
with respect to spin-2 gravitons by a loop factor in the amplitude.
Production through $W$ fusion is subleading.
However, for colliders 
running at energies not much above $M_Z$, it is possible that
external spin-0 KK production could impact collider observables.

The graviscalar coupling to heavy gauge bosons 
is most relevant for
$e^+e^-$ colliders working at energies not much larger than $M_Z$, as
in the case of LEP2. The inclusive differential
cross section for graviscalars accompanied
by a $Z$ boson is given by
\bea
\frac{d \sigma (e^+e^- \to ZH^{(\vec n )})}{dM_{\rm miss}}&=&\frac{G_F~M_Z^4 ~
\kappa^2~S_{\delta -1}~|1-6\hat\xi|^2~M_{\rm miss}^{\delta -1}}
{432 \sqrt{2}\pi ~s~ M_D^{\delta +2}}
\left( 1-4\sin^2\theta_W+8\sin^4\theta_W \right)\cr
&\times &
f\left (\frac{M_{\rm miss}^2}{s}, \frac{M_Z^2}{s}\right) , 
\label{cros}
\eea
where
\bea
\hat \xi & = & \frac{\xi M^2_{\rm miss}}{M^2_{\rm miss}-m^2_h
           +i\Gamma_h m_h},\label{csi} \\
f(x,y) & = &\frac{(1-x)^2+y(y-2x+10)}{(1-y)^2}\sqrt{(1-x)^2+y(y-2x-2)}, 
\eea
and
$M_{\rm miss}$ represents the graviscalar mass.  Because there is
a near-continous tower of KK masses for the graviscalars, 
a continuous distribution in $M_{\rm miss}$ results.

Notice that in the limit $m_h\to \infty$ the cross section becomes $\xi$ 
independent. In this limit the only scalars on the brane are the eaten
Goldstones. Then the result is consistent with the fact that no mixing
to the Ricci tensor can be written for Goldstone bosons (they couple
minimally to gravity). On the other hand,  for $m_h\ll M_{\rm miss}$
the cross section is proportional to $1-6\xi$. This corresponds to the 
high energy limit of a linear Higgs model, where the Higgs and the Goldstones
behave similarly. For $M_{\rm miss}\simeq m_h$ the above expression
is dominated by the production of a real Higgs that decays into invisible
graviscalars. This is the process we discussed in the previous section.

\begin{figure}[t]
\dofig{6in}{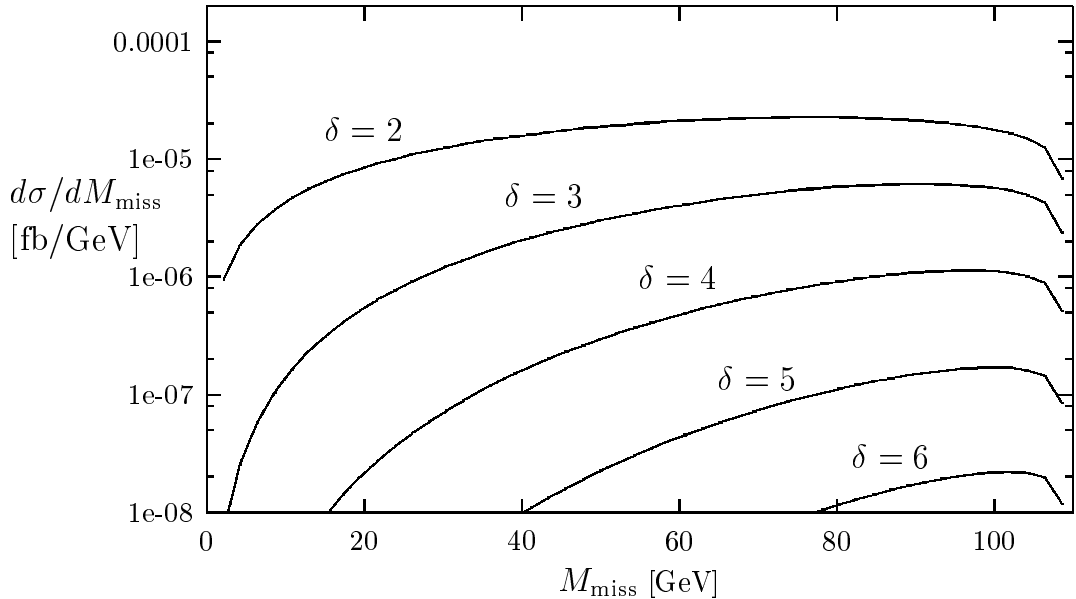} 
\caption{Differential signal cross-section 
of $e^+e^-\to Z\hn$ at LEP2 with $200\gev$
center of mass energy and for the parameter choices
$M_D=1\tev$ and $\xi=0$. $M_{\rm miss}$ is the missing
mass associated with escaping graviscalars. \label{dsigma}}
\end{figure}

In Fig.~\ref{dsigma} we plot the total differential cross-section
as a function of $M_{\rm miss}$ for LEP2 running at $\sqrt{s}=200\gev$
center of mass energy.  For this plot we have chosen
$M_D=1\tev$ and $\xi=0$.  For different choices of these parameters
one should rescale the curves by,
\beq
{\rm Rescale~factor~} = | 1-6\hat \xi |^2
  \left( \frac{1\tev}{M_D}\right)^{2+\delta}.
\eeq
The maximal integrated luminosity
at LEP2 running above $\sqrt{s}=200\gev$ is not expected to
exceed $1\xfb^{-1}$, summing over the four  
detectors.  Obtaining even a few {\rm total}
events of $e^+e^-\to Z\hn$ at LEP2 requires a large
rescaling factor. That is, $M_D$ needs to be substantially below
$1\tev$ or $\hat \xi$ must be substantially above 1 in order to produce
a few events.  Filling out a signal distribution in
the missing mass spectrum of $Z+{\rm invisible}$ events at LEP2 requires
an even higher enhancement factor. The expected background
for these events can be found in ref.~\cite{invisible Higgs}.
The missing
mass spectrum is peaked at $M_Z$, and the total selected event rate
with $Z\to  q\bar q$
corresponds to approximately 25 events in a 7 GeV mass bin
centered on $M_Z$ for $\sqrt{s}=189\gev$ 
(see the DELPHI article in ref.~\cite{invisible Higgs}).

In contrast, the production rate for spin-2 KK excitations 
in association with photons and Z is much higher for the
same parameter values~\cite{Giudice:1998ck}.  With a total of
$2\xfb^{-1}$, $M_D$ can be probed above $1.3\tev$
in the $\gamma + $ missing energy signature alone
for $\delta =2$. For these reasons
we conclude that KK graviscalar direct production at 
colliders is not likely to be as probing as KK graviton
direct production.
Nevertheless, the conclusions of the previous subsection
still hold. If the Higgs boson is kinematically accessible, there can
also be a resonant missing energy contribution coming from 
the graviscalar-Higgs
mixing, which is described by the Higgs invisible width in 
eq.~(\ref{inv width}).  This effect could be the {\it earliest} signal 
of extra dimensions at colliders.

\section{Graviscalars from non-factorizable geometries}

In this section we discuss the phenomenology of the radion in the
scenario of Randall-Sundrum~\cite{rs}. The extra dimensional space is now
a $S^1/Z_2$ orbifold parametrized by a coordinate $y \in [-\pi,\pi]$. The
geometry is the same as a slice of $AdS_5$
\beq
ds^2=e^{-2k r_c  |y|}\eta_{\mu\nu}dx^\mu dx^\nu-r_c^2 dy^2.
\eeq
where $1/k$ is the $AdS$ curvature radius and $r_c$ is the volume
radius.
The above metric solves Einstein's equations with a negative
bulk cosmological constant and in the presence of two branes
at $y=0$ and $y=\pi$ with respectively positive and negative tensions.
A field theory living on the $y=\pi$ brane experiences an exponential
red-shift 
$e^{-k r_c \pi }$ of all its mass parameters with respect to a theory living
at $y=0$. If one assumes that the SM lives at $y=\pi$, then it is enough
to have $k r_c\pi \sim 35$ to explain the hierarchy between 
$M_{\rm weak}/M_P$
in
a natural way. This fact motivates the great interest in this scenario.

The above metric admits two types of massless excitations
described by $\eta_{\mu\nu} \to g_{\mu\nu}(x)$, the usual 4-d  graviton,
and by $r_c\to T(x)$, the radion.  In terms of these two dynamical
fields, the metric can be recast in the form,
\beq
ds^2 =e^{-2k|y | T(x)} g_{\mu \nu} (x) dx^\mu dx^\nu -T^2(x)dy^2 .
\eeq
Of course in order to avoid violations of the equivalence principle
the modulus $T$ must acquire a mass. A mechanism that stabilizes
$T$ should also explain why its vacuum expectation value 
$\langle T(x) \rangle =r_c$ is somewhat larger 
than the $AdS$ radius: $ r_c \sim 11/k$.  Indeed Goldberger and Wise (GW)
\cite{Goldberger:1999uk} have
found a nice mechanism with these features. One consequence of the GW 
mechanism is that in order to have
$k \pi r_c\sim 35$  the radion should be somewhat lighter than 
the $J=2$ KK excitations. The radion is therefore likely to be 
the first state experimentally accessible in this scenario.

To write down the effective $3+1$ dimensional theory it is
 more convenient to express the radion field in terms of the field $\varphi$
defined by
\beq
\varphi \equiv \Lambda_\varphi e^{-k\pi (T-r_c)},~~~~
\Lambda_\varphi \equiv \langle \varphi \rangle = \sqrt{\frac{24M_5^3}{k}}
e^{-k\pi r_c},
\eeq
where $M_5$ is the Planck scale of the fundamental 5-dimensional theory. 
Integrating over the orbifold coordinate one then gets 
a canonically normalized effective 
action~\cite{Csaki:1999mp,Goldberger:1999un}
\beq
S_\varphi =\int d^4x \sqrt{-g} \left[ \frac{2M_5^3}{k}
\left( 1-\frac{\varphi^2}{\Lambda_\varphi^2}e^{-2k\pi r_c}\right) R
+\frac{1}{2} \partial_\mu \varphi \partial^\mu \varphi -V(\varphi )
+ \left( 1-\frac{\varphi}{\Lambda_\varphi}\right) T^\mu_\mu \right] .
\label{aktion}
\eeq
Here
$V(\varphi )$ is the potential which stabilizes the radion field
$\varphi$. For
phenomenological purposes, we are interested only in $T_\mu^\mu$
terms that are at most bilinear in the SM fields.
These are given by
\beq
{T^{(1)}}^\mu_\mu = 6\xi v \Dal h ,
\label{cazz1}
\eeq
\beq
{T^{(2)}}^\mu_\mu = (6\xi -1) \partial_\mu h\partial^\mu h +6\xi h \Dal h 
+2m_h^2h^2 + m_{ij}\bar\psi_i \psi_j -M_V^2 V_{A\mu}V_A^\mu .
\label{cazz2}
\eeq
Equations~(\ref{cazz1}) and (\ref{cazz2}) reduce to eq.~(\ref{newtrace})
after using the equations of motion for the matter fields, up to terms
containing three or more fields.

The existence of ${T^{(1)}}^\mu_\mu$ induces a kinetic mixing between $\varphi$
and $h$. After shifting $\varphi$ by its vacuum expectation value
$\langle \varphi \rangle =\Lambda_\varphi$,
the lagrangian containing bilinear terms in $\varphi$
and $h$ is given by
\beq
{\cal L} =-\frac{1}{2}  \varphi
\left( \Dal +m_\varphi^2 \right) \varphi
-\frac{1}{2} h \left( \Dal +m_h^2 \right) h -\frac{6\xi 
v}{\Lambda_\varphi} \varphi \Dal h .
\eeq
Here $m_\varphi$ is the mass parameter contained in $V(\varphi)$.
This lagrangian can be diagonalized by the field redefinitions
\beq
\label{deff1}
\varphi =\left( \sin \theta -\sin \rho \cos \theta \right) h'
+ \left( \cos \theta + \sin \rho \sin \theta \right) \varphi' ,
\eeq
\beq
\label{deff2}
h = \cos \rho \cos \theta ~h' - \cos \rho \sin \theta ~\varphi',
\eeq
\beq
\tan \rho \equiv \frac{6\xi v}{\Lambda_\varphi} ,~~~
\tan 2 \theta \equiv \frac{2\sin \rho ~m_\varphi^2}{\cos^2 \rho
(m_\varphi^2 -m_h^2)}.
\eeq
The new fields $\varphi'$
and $h'$ are mass eigenstates with eigenvalues 
\beq
m^2_{\varphi' , h'} = \frac{1}{2} \left[ (1+\sin^2 \rho ) m_\varphi^2
+\cos^2 \rho ~m_h^2 \pm \sqrt{ \cos^4 \rho (m_\varphi^2
-m_h^2)^2 + 4 \sin^2\rho ~m_\varphi^4} \right].
\label{ultim}
\eeq
Since we are dealing with an effective theory that contains
higher-dimensional operators suppressed by inverse powers of $\Lambda_\varphi$,
in the following we will often expand eqs.~(\ref{deff1})--(\ref{ultim}) 
and keep
only the leading terms in $1/\Lambda_\varphi$. Notice that, if 
$m_\varphi^2-m_h^2$ is small, one should retain higher orders in the
expression of the mixing angle $\theta$.

\subsection{Radion interactions with matter}

The interaction of the fields 
$\varphi$
and $h$ with fermions and massive gauge bosons is given by 
\beq
{\cal L} = -\frac{1}{v} \left( m_{ij}\bar\psi_i
\psi_j - M_V^2 V_{A\mu}V_A^\mu \right) \left[ h+ 
\frac{v}{\Lambda_\varphi}  \varphi
\right] .
\label{lagint}
\eeq
{}From this, we can obtain 
the interaction terms for the mass eigenstates $\varphi'$
and $h'$ by substituting eqs.~(\ref{deff1}) and (\ref{deff2}) into
eq.~(\ref{lagint}). We find that the radion decay widths into two fermions
and two massive gauge bosons are given by
\beq
\frac{\Gamma(\varphi'\to \bar ff,WW,ZZ)}{\Gamma(\hsm \to \bar ff,WW,ZZ)}
 =  \frac{v^2}{\Lambda^2_{\varphi}} (a_1-a_2)^2 ,
\eeq
\beq
a_1=\cos\theta+\sin\rho \sin\theta,~~~~{\rm and} ~~~~~~
     a_2 = \frac{\Lambda_\varphi}{v}\cos\rho\sin\theta .
\label{defa12}
\eeq
Here $\Gamma(\hsm \to \bar ff,WW,ZZ)$ are the usual
decay widths of the SM Higgs boson with $m_{\hsm}=m_{\varphi'}$.
Expanding at leading order in $1/\Lambda_\varphi$, we find
\beq
a_1-a_2= \left( 1-\frac{6\xi  m_\varphi^2}{
m_\varphi^2 -m_h^2}\right) +{\cal O}\left( \frac{1}{\Lambda_\varphi}\right) .
\label{blim}
\eeq
Notice that eq.~(\ref{blim}) 
vanishes in the conformal limit $m_h=0$, $\xi =1/6$.
The term proportional to $\xi$ in eq.~(\ref{blim}) can also be understood in a 
diagrammatic fashion as a Higgs-radion insertion on a Higgs-matter
coupling.

The coupling of the radion to two Higgs bosons is not 
model-independent because it can be affected by the stabilizing potential
$V(\varphi )$ after the field redefinitions in eqs.~(\ref{deff1}) and 
(\ref{deff2}). If we assume that the radion self-couplings in $V(\varphi )$
are small, then the decay width of the radion into two Higgs bosons is
determined by the interactions in eq.~(\ref{aktion}),
\beq
\label{rhh decay}
\Gamma (\varphi' \to h'h') =\frac{m_{\varphi'}^3}{32 \pi \Lambda_\varphi^2}
\left[
1-6\xi +2\frac{m_{h'}^2}{m_{\varphi'}^2} (1+6\xi) \right]^2 \sqrt{ 1
-4\frac{m_{h'}^2}{m_{\varphi'}^2}},
\eeq
at leading order in $1/\Lambda_\varphi$.
Again, the width vanishes in the conformal limit.

The production of both Higgs and radion fields at hadron colliders,
in the mass range of interest, is dominated
by gluon-gluon fusion. The effective vertex at momentum 
transfer $q$ along the scalar line is given by
\beq
\left[ {\varphi\over \Lambda_\varphi} 
b_3- \frac{1}{2}\left ({\varphi\over
\Lambda_\varphi}+{h\over v}\right ) F_{1/2}(\tau_t)\right]
\frac{\alpha_s}{8\pi}G_{\mu\nu}G^{\mu\nu},
\label{hgg}
\eeq
where $\tau_t=4m^2_t/q^2$, and $F_{1/2}$ is given in the
appendix.  We identify $q^2=m_{\varphi'}^2$ when we calculate
the on-shell decays of $\varphi'\to gg$.

The first term in eq.~(\ref{hgg}), 
where $b_3=7$ is the QCD $\beta$-function coefficient
in the SM, represents the QCD trace anomaly. The second term originates
from 1-loop diagrams involving virtual top quarks. The form factor
$F_{1/2}(\tau_t)$ is such that $F_{1/2}\to -4/3$ for $\tau\to\infty$,
and $F_{1/2}\to 0$ for $\tau\to 0$.
Notice that for $m_t^2\gg q^2$, the coupling to $\varphi$ becomes
proportional to the $\beta$-function for 5-flavors $b_3+2/3$,
consistent with top decoupling.  By substituting eqs.~(\ref{deff1}) 
and (\ref{deff2}) into eq.~(\ref{hgg}) we obtain
the $gg$ coupling to the mass eigenstates $h'$, $\varphi'$.
The decay width of the radion into two gluons is given by
\beq
\frac{\Gamma(\varphi'\to gg)}{\Gamma(\hsm \to gg)}
  =  \frac{v^2}{\Lambda^2_{\varphi}} 
    \frac{|2a_1b_3-(a_1-a_2)F_{1/2}(\tau_t)|^2}
    { |F_{1/2}(\tau_t)|^2} ,
\eeq
where $a_1$ and $a_2$ are given in eq.~(\ref{defa12}).

A similar expression describes the coupling to two photons
\beq
\left \{{\varphi\over \Lambda_\varphi} \left (b_{2}+b_Y\right)
-\left ({\varphi\over
\Lambda_\varphi}+{h\over v}\right )\left (F_1(\tau_W)+
\frac{4}{3}F_{1/2}(\tau_t)\right) \right  \}
\frac{\alpha_{EM}}{8\pi}F_{\mu\nu}F^{\mu\nu},
\label{hgaga}
\eeq
where $F_1(\tau_W)$ is a form factor from the loop with virtual $W$'s
(see appendix), while
$b_2=19/6$ and $b_Y=-41/6$ are the SM $SU(2)\times U(1)_Y$ $\beta$-funtion
coefficients. At $m_t^2/q^2\to \infty$ 
the coupling to $\varphi$ reduces to the
QED $\beta$-function with $e,\mu,\tau$ and $u,d,s,c,b$
\beq
b_2+b_Y-F_1(\infty)-\frac{4}{3}F_{1/2}(\infty)=-\frac{80}{9}=b_{\rm QED},
\eeq 
demonstrating again the correct decoupling behavior for heavy top
quark and $W$ boson. The decay width of the radion into two photons is
\beq
\frac{\Gamma(\varphi'\to \gamma\gamma)}{\Gamma(\hsm \to \gamma\gamma)}
 =  \frac{v^2}{\Lambda^2_{\varphi}} 
    \frac{|3a_1(b_Y+b_2)-(a_1-a_2)(4F_{1/2}(\tau_t)
              +3F_1(\tau_W))|^2}
    { |4F_{1/2}(\tau_t)+3F_{1}(\tau_W)|^2}.
\eeq

\subsection{Radion branching fractions and production}
\bigskip

In this
section we will compute the decay branching fractions and production 
cross-sections of the radion.
An accurate computation of the radion partial widths is best performed
by rescaling well-known SM Higgs boson partial 
widths~\cite{Djouadi:1998yw} according to the formulae
of the previous section.  
Since we will encounter regions of parameter
space where mixing between the radion $\varphi$ and Higgs boson $h$
is large, we must specify
which state we choose to call
the radion mass eigenstate $\varphi'$ and which state
we call the Higgs mass
eigenstate $h'$.
Our convention is to identify the radion mass eigenstate with the lighter
of the two solutions of eq.~(\ref{ultim}) if $m_\varphi < m_h$, and
the heavier of the two solutions if $m_\varphi > m_h$.  

\begin{figure}[t]
\dofigs{3.3in}{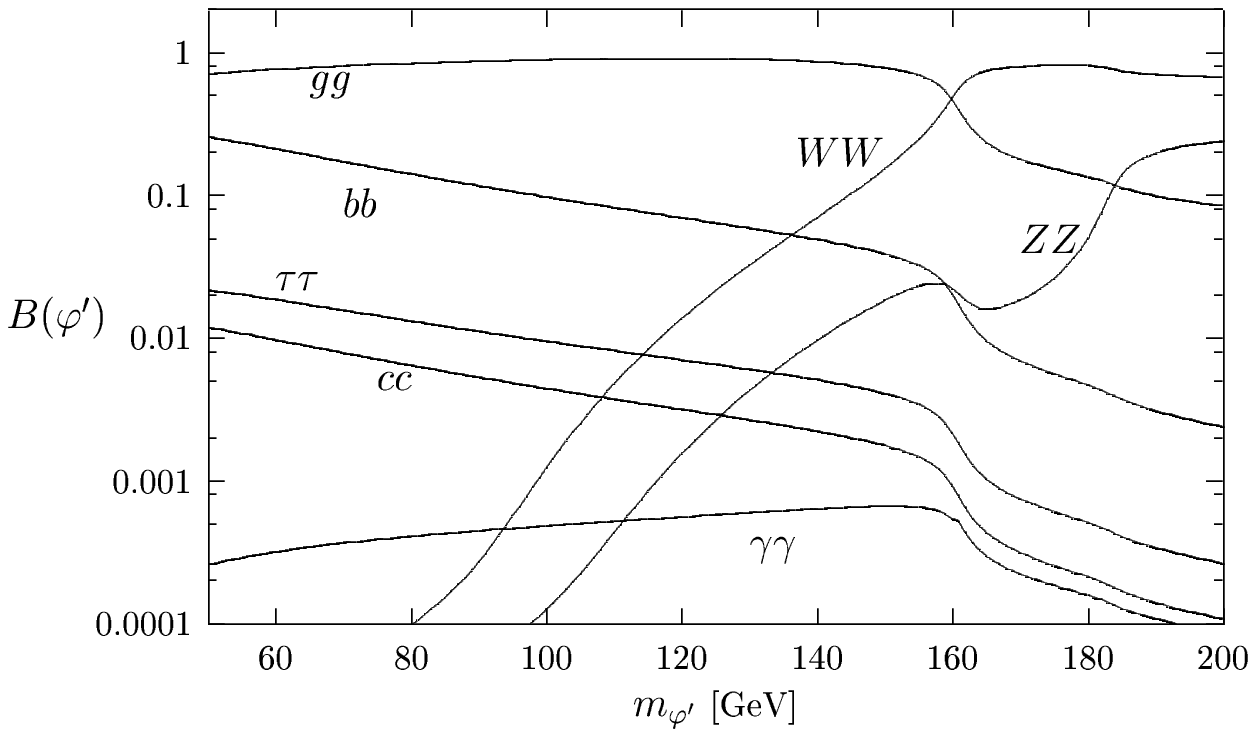}{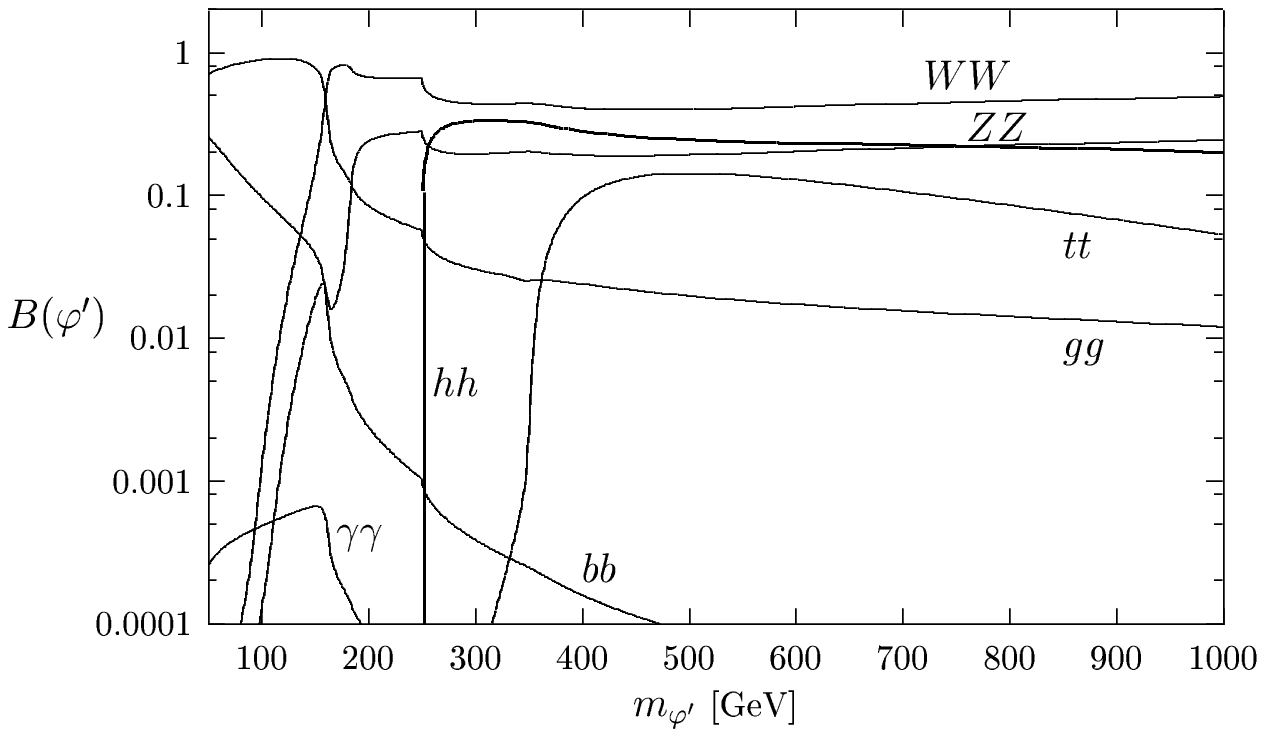}
\caption{Branching fractions of $\varphi'$ as a function
of its mass given $m_h=125\gev$, $\Lambda_\varphi =10\tev$ and
$\xi=0$.  The left and right panels are the same except a different
range in radion mass is covered. \label{rad1}}
\end{figure}

In Fig.~\ref{rad1} we plot the branching fractions of the radion
mass eigenstate as a function of its mass for
$\Lambda_{\varphi}=10\tev$, $m_h=125\gev$ and $\xi=0$.  Fig.~\ref{rad1}a 
shows the branching fraction over the light mass range of
$50\gev$ to $200\gev$.  Here, the branching fractions vary rapidly
over small changes in scale and many final states play a role in
the phenomenology of the radion.  Fig.~\ref{rad1}b plots the branching
fraction over a much wider mass range up to $1\tev$.  Additional states
become important at higher scales.  For example at $m_{\varphi'}>2m_t$
the top quark decay channel becomes accessible, and if
$m_{\varphi'}>2m_{h'}$ the $\varphi'\to h'h'$ decay becomes important.

The most important result of these two figures is the large branching
fraction into gluons for light radion mass.  
The branching fractions
into $b\bar b$ and two photons  --- the usual modes to search for the
light Higgs boson at colliders --- are suppressed 
in comparison to the SM Higgs boson.
At high $m_{\varphi'}$ we recover branching fractions
that are very similar to the SM.  This is because the one-loop
$\varphi'\to gg$ partial width starts to become overwhelmed by the $WW$, $ZZ$,
and $t\bar t$ partial widths.  Since the ratio of these latter partial
widths are the same for $\varphi'$ as for $h_{\rm sm}$ we recover the
SM branching ratios for these massive particles at high $m_{\varphi'}$.

\begin{figure}[t]
\dofigs{3.3in}{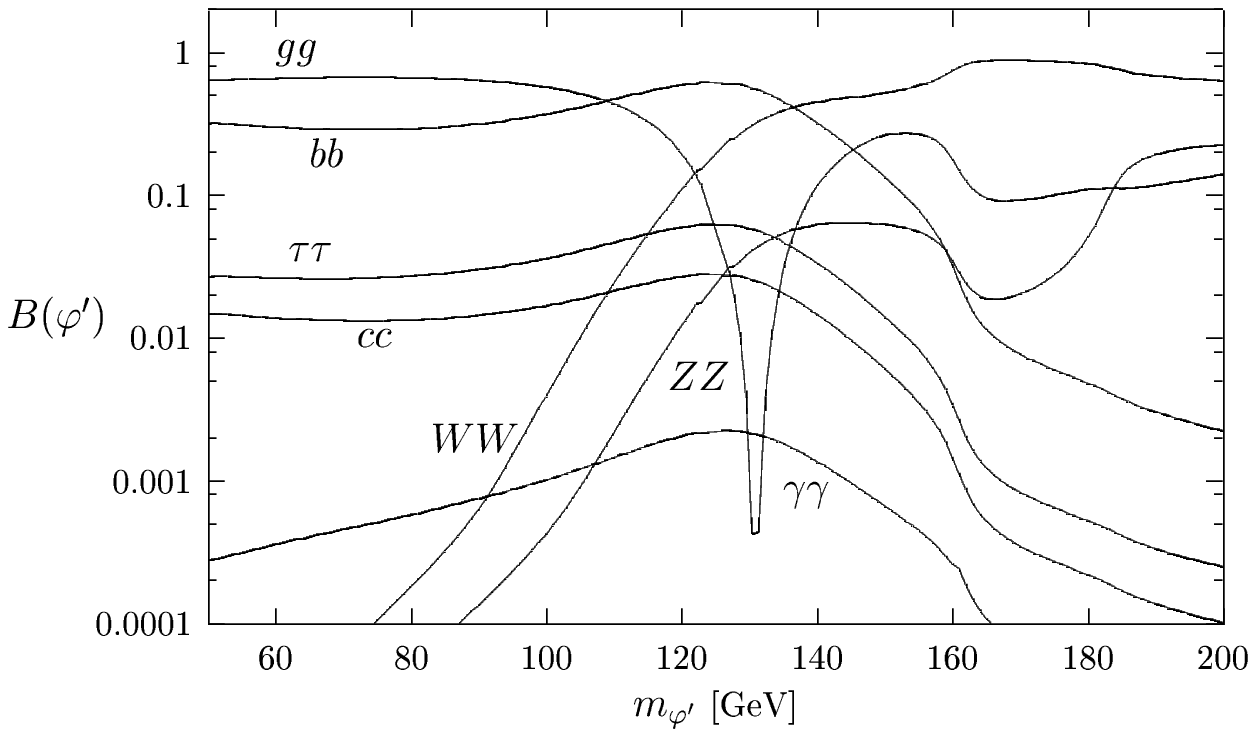}{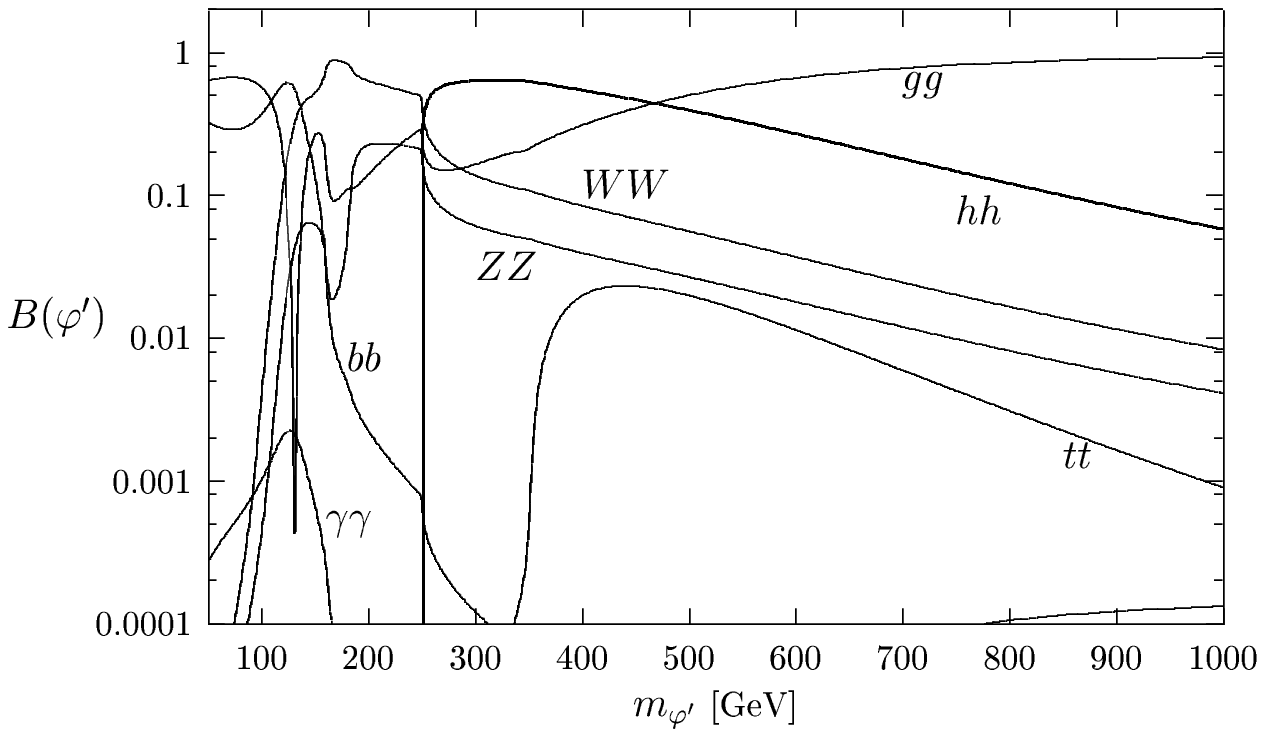}
\caption{Branching fractions of $\varphi'$ as a function
of its mass given $m_h=125\gev$, $\Lambda_\varphi =10\tev$ and
$\xi=1/6$.  The left and right panels are the same except a different
range in radion mass is covered. \label{rad2}}
\end{figure}

In Fig.~\ref{rad2} we construct the same branching fraction plots, except
this time we choose $\xi=1/6$.  For very light $\varphi'$ (mass less
than $80\gev$) the branching fractions are not much different than
what we obtained for $\xi=0$.  However, as we go higher in mass the
branching fraction of $b\bar b$ starts to climb and overtakes $gg$
for a radion with mass between
$110\gev$ to $140\gev$, and then it falls back down again rapidly.
The reason is because the radion mass eigenstate contains a heavy
mixture of the SM Higgs boson when its mass is near $m_h=125\gev$.
In the SM the $b\bar b$ partial width is always larger than $gg$ and
so it is not surprising that $\Gamma(b\bar b)>\Gamma(gg)$ when the
radion mixes heavily in the mass range
$m_{\varphi'}=125\pm 15\gev$.  We also see from the figure that 
$\Gamma (gg)$ falls rapidly at $m_{\varphi'}\simeq 130\gev$. This
is because the trace anomaly contribution cancels the one-loop
top quark contribution for this highly mixed state at that mass.

When $m_{\varphi'}$ gets very large, we see in Fig.~\ref{rad2}b that
the branching fraction into $gg$ becomes closer to $1$ again, while all
the others are dropping.  This is because when $m_{\varphi}\gg m_h$ 
and $\xi=1/6$ the couplings approach the conformal limit where
$a_1-a_2$ in eq.~(\ref{blim})
approaches zero. However, the radion coupling to gluons
does not approach zero in this limit because of the coupling to
the trace anomaly term.  The photon branching ratio is climbing
with the $gg$ branching ratio as it should since it also couples
to the trace anomaly, and it resurfaces on the plot in the lower right
corner.

The two cases $\xi=0$ and $\xi=1/6$ are somewhat special. For $\xi=0$, there is
no Higgs-radion mixing. For $\xi$ close to $1/6$, tree-level couplings 
of the radion to fermions and weak gauge bosons are suppressed and
$gg$ branching
fraction becomes dominant even for a very heavy radion.
Therefore, for a generic $\xi$ not too close to $1/6$,
the radion branching fraction 
phenomenology mostly follows what we found for $\xi=0$, except for the
region of large mixing, where our discussion of the $\xi=1/6$ case applies.

\begin{figure}[t]
\dofig{6in}{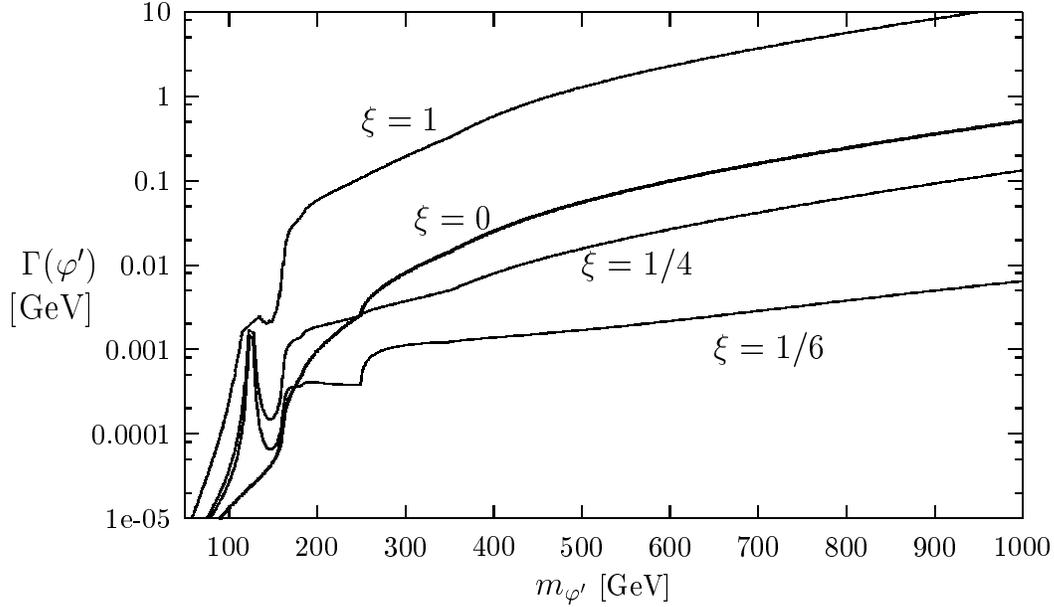}
\caption{The total width of the radion mass eigenstate given
$m_h=125\gev$ and $\Lambda_\varphi=10\tev$. The various
curves plotted correspond
to different values of the Higgs-curvature mixing parameter
$\xi$. \label{width}}
\end{figure}

The total width of the radion as a function of its mass is given
in Fig.~\ref{width} for both $\xi=0$ and $\xi=1/6$.  Since we
have chosen $\Lambda_{\varphi}=10\tev$ the radion is a very narrow resonance
scalar.  Near $m_{\varphi'}=125\gev$, the width of $\varphi'$ increases
significantly for the $\xi=1/6$ case.  Again, this is the region
where the radion is heavily mixed with the SM Higgs boson and so the
radion mass eigenstate can be thought of as ``half Higgs, half radion.''
The large overlap of the radion mass eigenstate with the SM Higgs boson
is what increases the width significantly in this region.
In Fig.~\ref{width} we also show the total
width of the radion for alternative choices of $\xi=1/4$ and $\xi=1$.
As expected, the widths are above that of $\xi=1/6$ and are rising
with mass, indicating the increasingly dominant contributions of
$WW$, $ZZ$ and $t\bar t$ to the total width.
Ratios of heavy radion branching fractions in
the detectable channels of $ZZ$, $WW$, and $t\bar t$
follow closely SM ratios in all cases but $\xi\simeq 1/6$.  
Overall production rates will depend
on $\Lambda_\varphi$ also, but they are straightforwardly  calculated from
the formulae given in this section.

The smallness of the radion width has its advantages and disadvantages
when one attempts to find evidence for this particle at a high energy
collider.  The disadvantage is that the production cross-sections are
closely correlated with the partial widths.  For example, 
\bea
\frac{\sigma (e^+e^-\to Z\varphi')}{\sigma(e^+e^-\to Z\hsm)}
 & = & \frac{\Gamma(\varphi' \to ZZ)}{\Gamma(\hsm\to ZZ)} \\
\frac{\sigma (q\bar q'\to W\varphi')}{\sigma(q\bar q'\to W\hsm)}
 & = & \frac{\Gamma(\varphi'\to WW)}{\Gamma(\hsm\to WW)} \\
\frac{\sigma (gg\to \varphi')}{\sigma(gg\to \hsm)}
 & = & \frac{\Gamma(\varphi'\to gg)}{\Gamma(\hsm\to gg)} .
\eea
Since the partial widths are reduced by overall factors of
$\sim v^2/\Lambda^2_{\varphi}$ with respect to the SM Higgs boson,
the production cross-sections are also lower.

On the other hand,
the small widths are an advantage when searching for invariant mass
peaks.  For example, $gg\to \hsm \to ZZ\to 4l$ would be much easier
if the Higgs boson had a narrower width.  The total width of the
SM Higgs boson is $\sim 70\gev$ for $m_{\hsm}=500\gev$ and climbs fast
at higher mass.  This is well above the 4 lepton invariant mass resolution
capabilities of the LHC detector.  If we estimate the invariant mass
resolution to be
\beq
\label{cazz3}
\frac{\Delta M_{4l}}{M_{4l}}=\frac{10\%}{\sqrt{M_{4l}(\gev)}}+0.005,
\eeq
we then get $\Delta M_{4l}=5\gev$ for $m_{\hsm} =500\gev$.  This
should be compared to $\Gamma(\hsm)=70\gev$ as we pointed out above.

The smaller width of the radion has the advantage of producing all
four-lepton events in a small invariant mass energy range.  The background
is then integrated over only this small range and the signal to background
ratio increases.  In the SM, where the heavy Higgs has a large width, all the
signal events occur over a much wider invariant mass energy range, and the
background must be integrated over this much larger range as well, reducing
the signal to background ratio.
The effective radion width is never smaller than the detector
resolution for $\Delta M_{4l}$, and so the background rate will be at least
as high as the integral over 
$m_{\varphi'}\pm \Gamma(\varphi')$.

As stated above,
the $\varphi'$ radion state has lower production cross-section and
lower partial widths than $\hsm$. We now attempt to investigate 
the signal significances of $gg\to \varphi' \to \gamma\gamma$
and $gg\to\varphi' \to ZZ$ by comparing them with the SM Higgs boson
signal significance.
We define signal significance as
\beq
S={\rm Significance}=\frac{\sigma_{\rm signal}}{\sqrt{\sigma_{\rm bkgd}}}
          \sqrt{{\cal L}^{-1}}
\eeq
where ${\cal L}^{-1}$ is the integrated luminosity. Significance greater
than 5 is considered a discovery.  Plots of SM significance curves in
this channel can be found in numerous places~\cite{lhc higgs}.

Neglecting low
luminosity and therefore low statistics possibilities, this definition
of significance allows us to make a direct comparison between the
SM Higgs boson and $\varphi'$, taking into account the total production
rate and the change in total width.  We therefore define
\beq
R^{4l}_S(\varphi') \equiv \frac{S(\varphi')}{S(\hsm)}
      = \frac{\Gamma(\varphi'\to gg)}{\Gamma(\hsm\to gg)}
         \frac{B(\varphi' \to ZZ)}{B(\hsm\to ZZ)}
 \sqrt{\frac{{\rm max} (\Gamma_{\rm tot}(\hsm), \Delta M_{4l})}
        {{\rm max} (\Gamma_{\rm tot}(\varphi'), \Delta M_{4l})}}
\eeq
as the ratio of the significance of the $\varphi'$ signal in the $4l$
channel compared to the significance of the $\hsm$ signal.
We also construct an analogous definition of $R^{\gamma\gamma}_S(\varphi')$.
Notice that $R(\varphi' )$ 
scales like only one power of
$1/\Lambda_\varphi$, allowing a significant
sensitivity to $\Lambda_\varphi$ substantially bigger than the 
weak scale.

\begin{figure}[t]
\dofigs{3.3in}{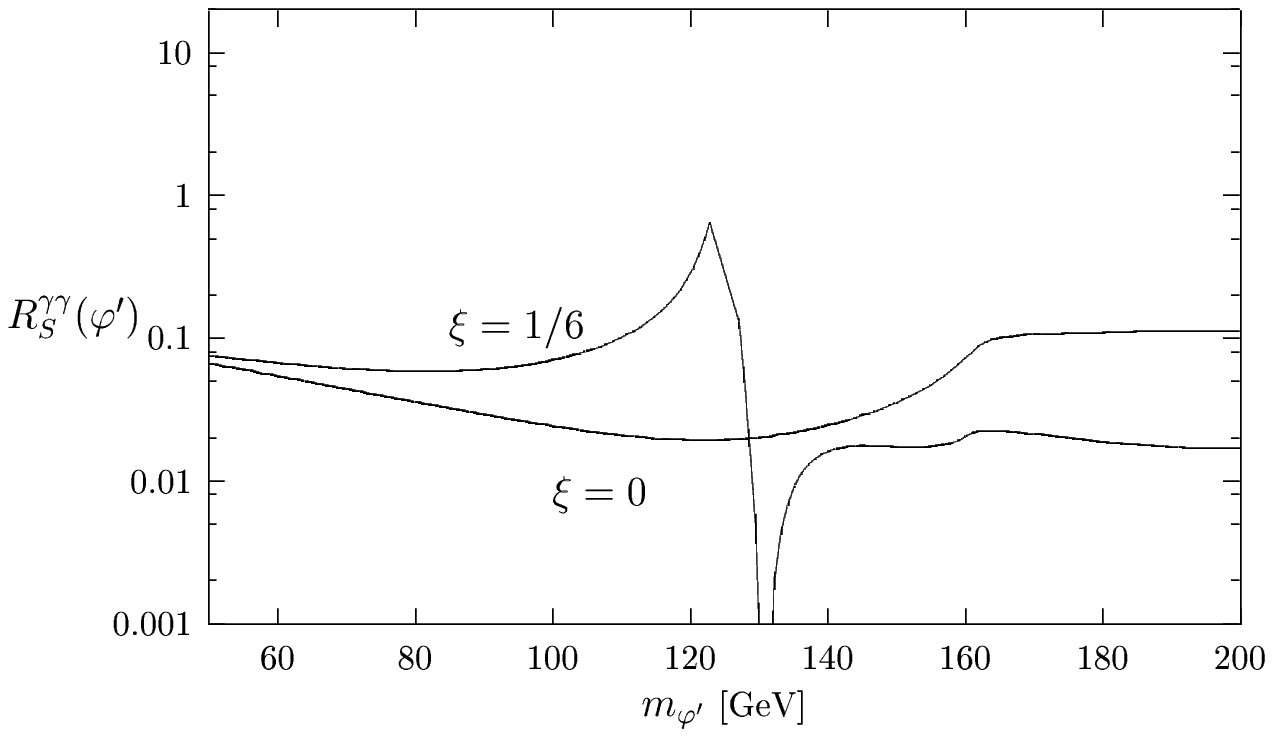}{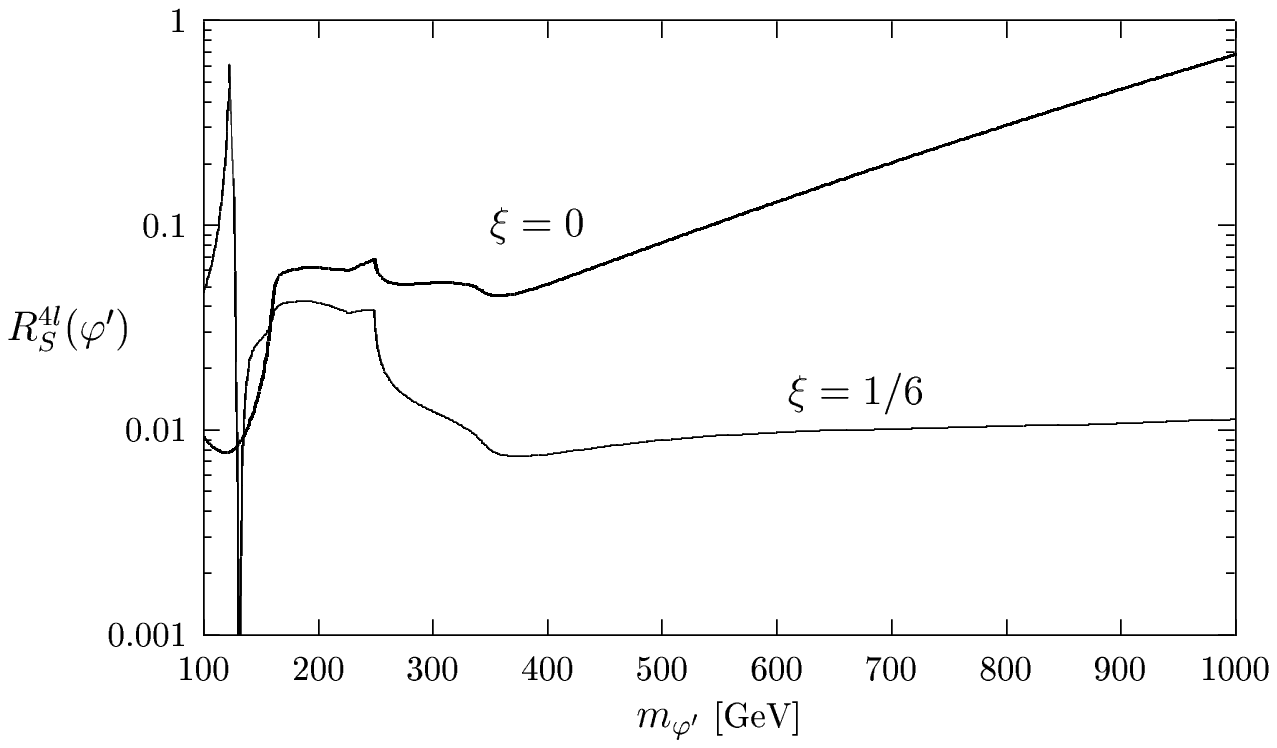}
\caption{Left: Plot of $R^{\gamma\gamma}_S(\varphi')$ as a function
of the radion mass for $m_h=125\gev$ and $\Lambda_\varphi=10\tev$.
$R^{\gamma\gamma}_S(\varphi')$ is the ratio of the signal significance
of the radion in the $gg\to \varphi'\to\gamma\gamma$ channel 
compared to that of the SM Higgs boson with the same mass.
The two lines correspond to different choices of $\xi$, the 
curvature-Higgs mixing parameter.  Right: This plot is the same as the
left panel plot except the y-axis is $R^{4l}_S(\varphi')$, which
represents the ratio of the signal significance of the radion in the
$gg\to \varphi'\to Z^{(*)}Z^{(*)}\to 4l$ channel compared to that of the SM
Higgs boson with the same mass. \label{hyy}}
\end{figure}

In Fig.~\ref{hyy} we plot  $R^{\gamma\gamma}_S(\varphi')$
and  $R^{4l}_S(\varphi')$ as a function of the radion mass.
Fig.~\ref{hyy}a covers the low mass range where the
SM significance is respectable at the CERN LHC running
at $14\tev$ center of mass energy.  
The significance of the $\gamma\gamma$ signal usually is between
$1/10$th to $1/100$th of the SM Higgs boson
significance.  This particular set of parameters
is therefore not detectable at the LHC with less than $10\xfb^{-1}$ since
the significance of the SM signal is never above $50$ for this integrated
luminosity.  However, there is a small region 
near $m_{\varphi'}\simeq 125\gev$ for $\xi=1/6$ where the
significance peaks and then drops.  
This is the heavily mixed region where the radion
mass eigenstate is ``half Higgs, half radion.''  
Therefore, we expect the radion significance to approach
about $1/2$ the SM significance in this narrow region, then drop
quickly at $m_{\varphi'}\simeq 130\gev$
due to the cancellation between the trace anomaly
term and the one-loop term in $\Gamma(gg)$, which sets the
production cross-section. 

The $gg\to \varphi'\to ZZ\to 4l$ signal shows an interesting
dependence on $\xi$.  First, at $125\gev$ and $\xi=1/6$ we see a peak
of about $1/2$ for the same reason it occured in the $\gamma\gamma$ case.
At much higher radion mass we see that the $\xi=1/6$ curve is falling
slightly and remains near $1/100$ the significance of the SM.  
On the other hand,
the $\xi=0$ curve shows a steady rise in significance.  This is due
to the branching fractions becoming more like those of the SM and because
the ratio $\sigma(gg\to \varphi')/\sigma(gg\to \hsm)$ 
rises with energy.  Furthermore, 
the total width of $\hsm$ is becoming large at these high masses, and
the background increases dramatically in a $\pm \Gamma(\hsm)$ bin around 
$m_{\hsm}$, whereas the radion width is much smaller and so the
background within the $m_{\varphi'}\pm \Gamma(\varphi')$ bin
is significantly smaller.  
Therefore, at very high mass, the significance in detecting
the radion can rise to nearly that of the SM even for 
$\Lambda_{\varphi}=10\tev$. The sensitivity 
for  a generic value of $\xi$ different from $1/6$ and
$0$ is essentially similar to the case $\xi=0$, 
apart from the region of large mixing where it peaks 
and dips like in the $\xi=1/6$
case. 
One general conclusion is that a heavy radion has good chances of being 
detected, except for $\xi$ very close to  $1/6$.

Finally, we estimate the reach LHC has to discover the
radion.  The parameters determining the radion phenomenology
are $m_{\varphi'}$, 
$\Lambda_\varphi$, $\xi$, and $m_{h'}$.
The parameters $m_{\varphi'}$ and 
$\Lambda_\varphi$ are perhaps the most important, 
since $m_{\varphi'}$ largely determines
the search strategy ($2\gamma$ or $4l$ searches), while
$\Lambda_\varphi$ sets the overall production rate.  We therefore
set $\xi=0$ and $m_{h}=125\gev$ fixed, and analyze
detection prospects as a function of radion mass and
radion coupling to $T^\mu_\mu$.

\begin{figure}[t]
\dofigs{3.3in}{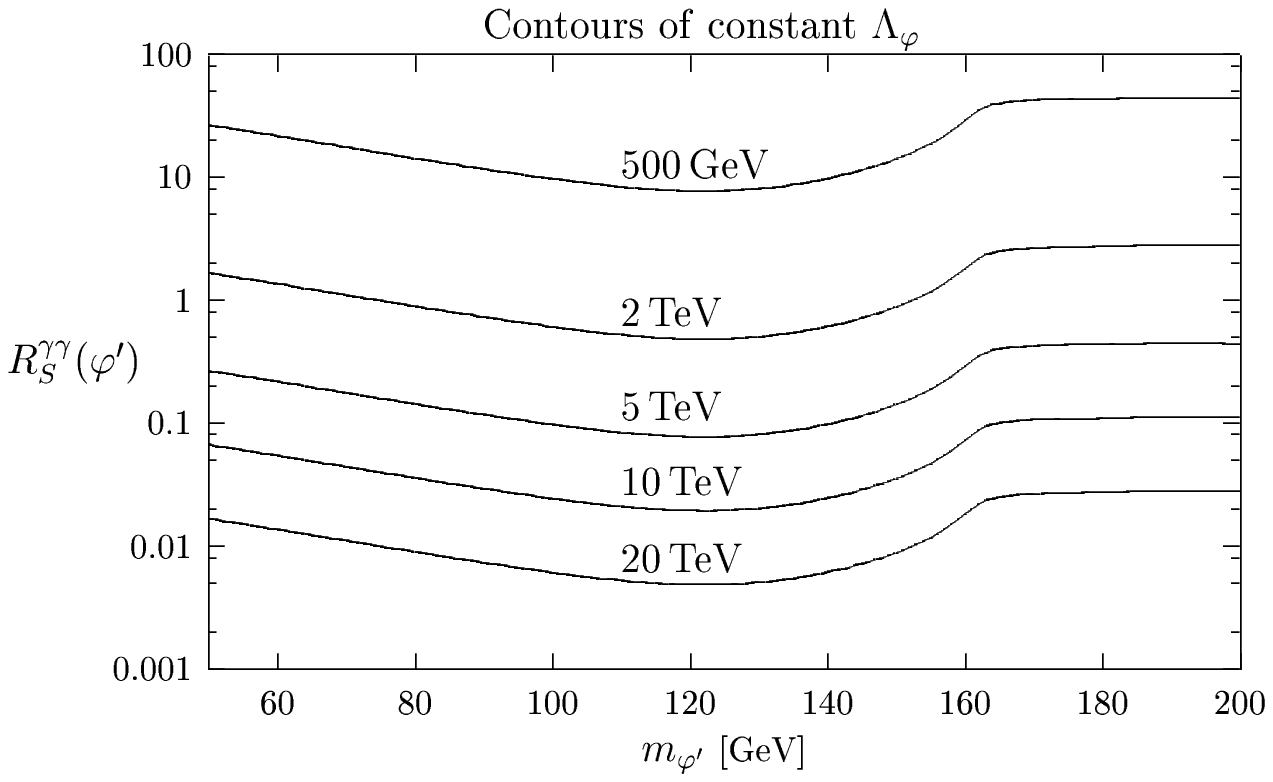}{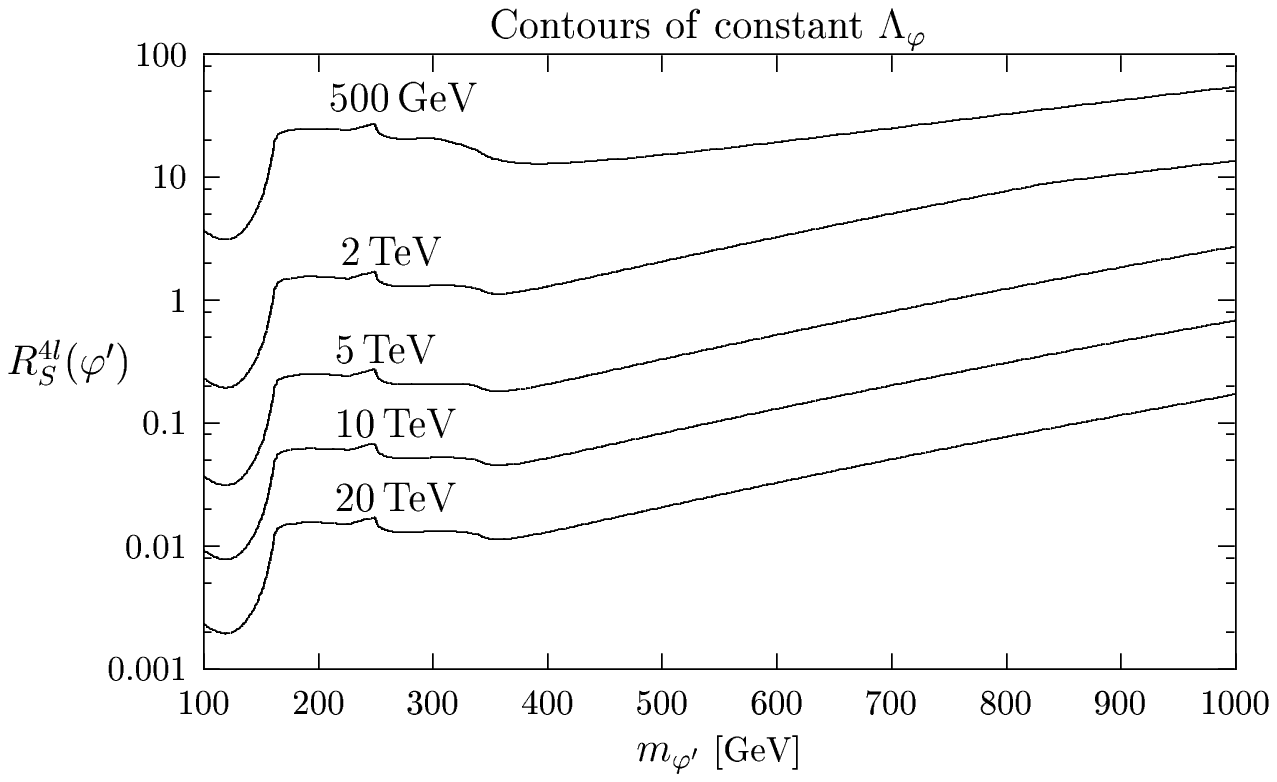}
\caption{Plots of the ratio of radion signal significance to the
signal significance of a SM Higgs boson of the same mass in the
$\gamma\gamma$ and $4l$ channels.  These plots are made
for $\xi=0$ and $m_{h'}=125\gev$. The different lines on the
graph represent various choices of constant $\Lambda_\varphi$.
\label{vary lambda}}
\end{figure}

In Fig.~\ref{vary lambda} we plot $R^{\gamma\gamma}_S(\varphi')$
and $R^{4l}_S(\varphi')$ as a function of radion mass for various
choices of $\Lambda_\varphi$.  We estimate search reach in
$\Lambda_\varphi$ by requiring the signal significance to
be greater than or equal to that of the SM: $R(\varphi' )\geq 1$.  We must
be careful to identify the correct signal over the mass region
which allows discovery.  For example, the SM Higgs boson can be
found in the $2\gamma$ channel in the mass range 
$m_{h_{\rm sm}}\lsim 150\gev$.  Beyond $150\gev$, the two photon signal is not
a useful strategy to search for the SM Higgs boson.  In 
Fig.~\ref{vary lambda}a we see that $\Lambda_\varphi\lsim 2\tev$
allows for greater radion signal significance than the SM in
the mass range $m_{\varphi'}\lsim 150\gev$.  

Likewise, the $4l$ signal is effective for the SM over the rest of
the mass range, provided enough luminosity is attained ($\sim
100\xfb^{-1}$).  In Fig.~\ref{vary lambda}b we plot the ratio
of the signal significance of the radion to that of the SM Higgs
boson in the $4l$ channel.  If we define the $\Lambda_\varphi$ mass reach as
the value of $\Lambda_\varphi$ for which $R^{4l}_S(\varphi')=1$, we see
that it can vary depending on $m_{\varphi'}$.  For lower values of
$m_{\varphi'}$ it appears that $\Lambda_{\varphi}\lsim 2\tev$ can
be probed rather effectively, whereas for higher $m_{\varphi'}\gsim 500\gev$,
$\Lambda_\varphi$ appears to be probed at the LHC up to $5\tev$.

The above discussion represents a first estimate on the mass reach of
$\Lambda_\varphi$ at the LHC.  An obvious refinement of this analysis
is to realize that the SM signal significance varies considerably
over the range of SM Higgs boson masses. For example, for Higgs mass
between $200\gev$ and $400\gev$, the SM signal significance is
greater than 15 with $100\xfb^{-1}$ at the LHC running at
$\sqrt{s}=14\tev$~\cite{lhc higgs}.  
Therefore, in order for the radion to have a significance
greater than 5 necessary for discovery, $R^{4l}_S(\varphi')$ need
only be greater than $1/3$.  This raises the probing capability of
$\Lambda_\varphi$ beyond $3\tev$ in this region.

We have convoluted the SM significance~\cite{lhc higgs}
with the radion significance calculations.  We then have estimated
the search capability of the radion at the LHC with $100\xfb^{-1}$
of data and find,
\bea
110\gev < m_{\varphi'}< 150\gev   & 
         \Rightarrow & 2\tev \lsim \Lambda_\varphi\lsim 3\tev \nonumber \\
150\gev < m_{\varphi'}< 550\gev & \Rightarrow & 
                        3\tev\lsim \Lambda_\varphi\lsim 7\tev \nonumber \\ 
550\gev < m_{\varphi'}\lsim 950\gev & \Rightarrow & 
             7\tev\gsim\Lambda_\varphi\gsim 4\tev . \nonumber
\eea
We do not go above $950\gev$ since no realiable SM computation exists
in this region to compare to.  One can of course recast the analysis
into an estimate of the $m_{\varphi'}$ mass range that could be discovered
given a fixed $\Lambda_{\varphi}$.  Either way, the results here indicate
that the LHC can effectively search for evidence of non-factorizable
geometry in the multi-TeV region --- an important mass range if
these ideas have some relevance to the hierarchy problem.

\section{Conclusions}

{}From a four dimensional point of view,
gravity in higher dimensional spaces implies the existence of many
new states of spin 2, 1 and 0  interacting with SM particles.  The
existence
of these particles can have an important impact on cosmological,
astrophysical and collider observables.  For example, massive spin-2
Kaluza-Klein excitations of the graviton can lead to detectable 
missing energy signatures at high-energy colliders.
In this article we pointed out the importance of considering
graviscalars 
in the phenomenological implications of higher-dimensional metrics.
By general covariance, the graviscalars that couple to SM states
must do so through the trace of the energy-momentum tensor 
$T_\mu^\mu$. Our first remark is that in the SM, as in any theory with 
fundamental scalar particles, $T_\mu^\mu$ already at the two derivative
level admits the introduction of a new dimensionless parameter $\xi$.
This parameter corresponds to a lagrangian operator mixing the SM Higgs
doublet bilinear with the Ricci scalar of the induced metric.
While $\xi$ does not affect the coupling of $J=2$ gravitons on-shell,
it is crucial for discussing the phenomenology of the spin zero
gravitons.
We considered two interesting scenarios with low scale quantum
gravity,
the case of $\delta \geq 2$ large and flat extra-dimensions (ADD) and
the case
of just one warped new dimension (RS).

In the ADD case, the cross section for producing $J=2$ KK states grows 
 like $(E/M_D)^{2+\delta}$ with respect to SM backgrounds. Therefore
the higher the center of mass energy the better the sensitivity. However,
the production of $J=0$ KK is in most cases suppressed either
by the anomaly loop factor or by an additional
power of $m_Z^2/E^2$ relative to $J=2$,
and correspondingly leads to a lower sensitivity at high energy.
This fact is a simple consequence of the relation between $T_\mu^\mu$
and the parameters that explicitly break scale invariance, as shown
in eq.~(\ref{newtrace}). There are however important exceptions to the
subdominance of
$J=0$ to $J=2$. One is represented  by processes involving longitudinal
massive vector bosons which, for $\xi\not= 1/6$,  scale in energy just
like the $J=2$ case. The underlying reason for this is that, in high-energy 
processes, the longitudinally polarized vector bosons 
are equivalent to the eaten Goldstone bosons. Another important 
exception is the mixing arising for $\xi\not=0$ between
the SM Higgs boson and the graviscalars, which is not forbidden by
any symmetry.
Due to the huge number ${\cal O} (M_P^2)$
of closely spaced KK level almost degenerate in mass with the Higgs, 
the physical consequence is a Higgs  invisible decay into  
states with {\it just one} graviton. For $m_h< 2M_W$ the 
relative size of this invisible width compared  to the standard one
 into $b\bar b$  is roughly $8\pi ^2v^2m_h^\delta /(\lambda_b^2 
M_D^{\delta +2})$. 
So while this effect cannot enjoy the $(E/M_D)^{2+\delta}$ growth
of the $J=2$ signals, it is made more relevant by the smallnes of
$\lambda_b$.
We have computed this invisible decay rate
and demonstrated that it can lead to a branching fraction near 1 for
reasonable parameters, greatly impacting Higgs boson search strategies
at LEP2, Tevatron and LHC.

In the Randall-Sundrum scenario, with one extra-dimension, the only
 graviscalar is the radion $\varphi$ .  It is coupled with 
$1/\Lambda_\varphi\sim 1/{\rm TeV}$ strength
to the trace of the energy-momentum tensor.  
This is qualitatively different than
the ADD case, where a dense tower of spin-0 KK states couple with
$1/M_P$ strength.  If the Ricci-Higgs mixing $\xi$ is zero, then the radion
exists as an independent state from the SM Higgs boson.  SM Higgs boson 
phenomenology is unaffected.  However, the radion can be produced at high 
rate in $gg$ fusion through the QCD trace anomaly
and can be discovered through its different
decay modes to SM fermions, vector bosons,  and even SM Higgs bosons if the
radion mass is large enough.   On general grounds, however, we expect
$\xi\not = 0$
in the action, so that the radion and the SM Higgs boson mix into two
new mass eigenstates, neither one acting entirely like the SM Higgs boson
or the naive radion.  Branching fractions into SM states can be quite
different
depending on this coupling.  For example, we demonstrated in the
previous
sections that $\varphi'$ decays are qualitatively altered by choosing
$\xi=0$ (minimally coupled) or $\xi=1/6$ (conformally coupled).
Most notably, the decay branching fractions to $WW$ and $ZZ$ at high
masses, $m_{\varphi'}>500\gev$, can be a factor of 50 different for these
two choices of $\xi$ given all other parameters the same. 
We have studied the search capability at the LHC as a function of the radion
mass and coupling.  One welcome fact in hadron collisions is that the
radion coupling to two gluons does not vanish for a heavy radion.
This is due to the anomalous origin of this coupling and should be contrasted
to the top-loop induced Higgs coupling which diminishes for a heavy Higgs.
We find that the LHC can probe the fundamental scale $\Lambda_\varphi$ in
the multi-TeV region for practically all values of $\xi$ and for 
$m_{\varphi'}< 1\tev$. We conclude that the LHC can effectively test 
the relevance of non-factorizable geometries to the hierarchy problem.

\bigskip
\noindent
{\it Acknowledgements:}  
R.R. and J.D.W. wish to thank the ITP Santa Barbara for 
its support during part of this work (NSF Grant No.\ PHY94-07194).
We thank the Aspen Center for Physics, where this work
was initiated. R.R. thanks Vincenzo Napolano for useful
conversations.  We also thank
A.~Blondel, M.~Kado and P.~Janot for providing us with some useful
information. 


\section*{Appendix}

In the coupling of $\varphi'$ to $gg$ and $\gamma\gamma$ we encountered
the form factors, $F_{1/2}(\tau_t)$ and $F_1(\tau_t)$,
where $\tau_t=4m_t^2/q^2$ and $\tau_W=4m^2_W/q^2$.  We utilize the
notation of Ref.~\cite{Gunion:1989we} in the definition of these functions:
\bea
F_{1/2}(\tau) & = & -2\tau [1+(1-\tau)f(\tau)] \\
F_1(\tau) &= & 2+3\tau+3\tau(2-\tau)f(\tau),~~~{\rm where} 
\eea
\beq
f(\tau)  =  \left\{ \begin{array}{cc}
      \left[ \sin^{-1}\left( 1/\sqrt{\tau}\right)\right]^2, &
               {\rm if}~\tau\ge 1, \\
      -\frac{1}{4}\left[ \ln (\eta_+/\eta_-)-i\pi\right]^2, &
               {\rm if}~\tau < 1,
                      \end{array}  \right.
\eeq
and
\beq
\eta_{\pm} = 1 \pm \sqrt{1-\tau} .
\eeq
The functions have the following limits,
\beq
F_{1/2}(\tau) = -\frac{4}{3}~~~{\rm and}~~~F_1(\tau)=7~~~~
   {\rm as}~\tau\to\infty .
\eeq


\end{document}